\documentclass[conference]{IEEEtran}
\usepackage{cite}
\usepackage{soul}
\usepackage{amsmath,amssymb,amsfonts}
\usepackage{algorithmic}
\usepackage{graphicx}
\usepackage[table,xcdraw]{xcolor}  
\usepackage{comment}
\usepackage{longtable}
\usepackage{booktabs} 
\usepackage{array, makecell} 
\usepackage{geometry} 
\usepackage{amssymb} 
\usepackage{enumitem}
\usepackage{hyperref}
\usepackage{url} 
\usepackage{tabularx}  

\usepackage{caption}
\definecolor{Gray}{gray}{0.85}
\usepackage{lipsum}
\usepackage{placeins}

\usepackage[textsize=scriptsize]{todonotes}
\newcommand{\ignore}[1]{}
\newcommand{\cmark}{\ding{51}}
\newcommand{\xmark}{\ding{55}}
\usepackage{pifont} 
\usepackage{multirow} 

\usepackage{float}

\def\BibTeX{{\rm B\kern-.05em{\sc i\kern-.025em b}\kern-.08em
    T\kern-.1667em\lower.7ex\hbox{E}\kern-.125emX}}
\begin{document}

\title{Deep Learning Based XIoT Malware Analysis: A Comprehensive Survey, Taxonomy, and Research Challenges}

\author{
    Rami Darwish\IEEEauthorrefmark{1}, Mahmoud Abdelsalam\IEEEauthorrefmark{2}, Sajad Khorsandroo\IEEEauthorrefmark{3}\\
    \IEEEauthorblockA{Department of Computer Science, North Carolina A\&T State University, Greensboro, NC, USA}
    \\
    Email: \IEEEauthorrefmark{1}rhdarwish@aggies.ncat.edu, \IEEEauthorrefmark{2}mabdelsalam1@ncat.edu, \IEEEauthorrefmark{3}skhorsandroo@ncat.edu
}

\maketitle

\begin{abstract}
The Internet of Things (IoT) is one of the fastest-growing computing industries. By the end of 2027, more than 29 billion devices are expected to be connected. These smart devices can communicate with each other with and without human intervention. This rapid growth has led to the emergence of new types of malware. However, traditional malware detection methods, such as signature-based and heuristic-based techniques, are becoming increasingly ineffective against these new types of malware. Therefore, it has become indispensable to find practical solutions for detecting IoT malware. Machine Learning (ML) and Deep Learning (DL) approaches have proven effective in dealing with these new IoT malware variants, exhibiting high detection rates. In this paper, we bridge the gap in research between the IoT malware analysis and the wide adoption of deep learning in tackling the problems in this domain. As such, we provide a comprehensive review on deep learning based malware analysis across various categories of the IoT domain (i.e. Extended Internet of Things (XIoT)), including Industrial IoT (IIoT), Internet of Medical Things (IoMT), Internet of Vehicles (IoV), and Internet of Battlefield Things (IoBT).

\end{abstract}


\section{\textbf{Introduction}}
\label{sec:introduction}
In the era of ubiquitous computing and IoT, the broad concept of 'XIoT' encompasses specialized domains—each tailored to specific operational needs and technological environments. XIoT includes familiar domains such as IoT, IIoT, IoMT, IoV, and IoBT. Together, these form the backbone of what is now known as the Extended Internet of Things (XIoT).

The XIoT domain primarily enhances consumer usability and lifestyle through smart devices in personal and home environments, setting the stage for more sophisticated applications across its other specialized counterparts. For instance, the IIoT builds upon essential IoT technologies to optimize industrial processes through advanced automation and real-time data analytics \cite{turner2021human}. Similarly, the IoMT is transforming healthcare by integrating intelligent medical devices that improve diagnostic accuracy and patient care \cite{manickam2022artificial}. Furthermore, the IoV enhances transportation systems by enabling vehicle-to-vehicle (V2V) and vehicle-to-infrastructure (V2I) communication, thereby increasing road safety and improving traffic management \cite{sadiku2021overview}. On the military front, the IoBT incorporates IoT technologies to enhance tactical operations through better situational awareness and automated battlefield support \cite{russell2018internet}. 
All of such have particular integration that facilitates seamless connectivity and enables more intelligent interactions between devices.

Despite the advancements in these domains, they also introduce significant security vulnerabilities, evidenced by the rise in malware attacks such as Mirai and Gafgyt \cite{website_IoT_malware_loss}. These attacks have notably devastated the XIoT domain, leading to extensive service disruptions and financial losses, with the global impact quantified in the billions \cite{website_IoT_malware_loss}. Traditional malware detection methods have struggled to tackle sophisticated threats effectively. Conventional defenses, often reliant on outdated signature-based detection, are increasingly bypassed by attackers \cite{okoli2024machine}. In contrast, machine learning techniques have shown promise in identifying malware through pattern recognition, although they require significant data analysis and reverse engineering efforts.

The 2018 WannaCry ransomware attack on Taiwan Semiconductor Manufacturing Company (TSMC), the world’s largest semiconductor manufacturer, underscores the critical vulnerabilities inherent in IIoT environments. This attack, which led to a significant production shutdown, was initiated through a USB device infected with a variant of the WannaCry ransomware. The incident highlights the severe risks posed by physical attack vectors, such as USB devices, within IIoT settings—where the convergence of digital and physical systems can result in extensive operational disruptions \cite{kim2022iiot}.

Similarly, IoMT domain faces unique cybersecurity challenges due to the sensitive nature of healthcare data and the critical importance of medical device functionality. Cyberattacks on IoMT systems often involve data breaches, man-in-the-middle assaults, probe attacks, decryption of network communications, and DoS attacks, all of which can lead to significant privacy and security issues. Notably, an IBM survey reveals that healthcare organizations have incurred the largest financial losses due to data theft, further emphasizing the vulnerability of IoMT environments. A striking example of this is the recent infiltration of Singapore's health system, where hackers accessed the personal information of 1.5 million patients, including the outpatient prescription data of 160,000 individuals, among whom was Singapore's Prime Minister \cite{saheed2021efficient}.

These examples from the IIoT and IoMT domains illustrate the broader risks facing the entire XIoT. As these domains become more interconnected and integral to critical infrastructure, the consequences of security breaches become increasingly severe. This underscores the urgent need for advanced, domain-specific defenses capable of addressing the unique threats posed to each area of the XIoT.

In light of these issues, this survey aims to complement existing works and provide insights into the critical deep learning-based techniques used to analyze malware in XIoT, covering the latest developments. Deep learning significantly reduces manual labor and improves detection accuracy by automatically extracting complex patterns and learning from vast datasets.

\subsection{\textbf{Prior Surveys and Limitations}}
\begin{table}[!t]
\centering
\caption{Comparison of recent surveys on DL-based XIoT Malware Analysis and their comprehensive coverage of XIoT domains. Key: \cmark = Comprehensive coverage based on numerous studies; \xmark = Topic not covered; \cmark(M) = Minimal coverage based on a few studies. }
\label{dl-based-analysis}
\setlength\tabcolsep{4.5pt}
\begin{tabular}{@{}lccccccc@{}}

\toprule
\textbf{Survey} & \textbf{IIoT} & \textbf{IoMT} & \textbf{IoV} & \textbf{IoBT} & \textbf{IoT} \\ \midrule
Madan et al (2022). \cite{madan2022tools}
                                    & \xmark        & \xmark        & \xmark       & \xmark        & \cmark(M)                        \\
Gopinath et al (2023). \cite{gopinath2023comprehensive}                                       & \cmark(M)        & \xmark        & \xmark       & \cmark        & \cmark                       \\
Gaurav et al (2023). \cite{gaurav2023comprehensive}                                       & \xmark        & \xmark        & \xmark       & \cmark        & \cmark(M)                     \\
Ngo et al (2020). \cite{ngo2020survey}                     & \xmark        & \xmark        & \xmark       & \cmark        & \cmark(M)                        \\
Aslan et al (2020). \cite{aslan2020comprehensive}                                & \xmark        & \xmark        & \xmark       & \xmark        & \cmark(M)                     \\
Al-Garadi et al (2020). \cite{al2020survey}                                    & \xmark        & \xmark        & \xmark       & \xmark        & \cmark(M)                        \\
Raju et al (2021). \cite{raju2021survey}                                  & \xmark        & \xmark        & \xmark       & \xmark        & \cmark                    \\
Wazid et al (2019). \cite{wazid2019iomt}                             & \xmark        & \xmark        & \xmark       & \xmark        & \cmark(M)                      \\
Our Survey                             & \cmark        & \cmark        & \cmark       & \cmark        & \cmark                    \\ \bottomrule
\end{tabular}

\end{table}

In this subsection, we provide a concise overview of research surveys that analyze malware in XIoT domains using deep learning methods. The comparisons of these studies are illustrated in Table~\ref{dl-based-analysis}. 

Several surveys ~\cite{madan2022tools, gaurav2023comprehensive} has focused on the utilization of ML in the IoT domain.
Madan et al.~\cite{madan2022tools} provides a comprehensive review of the methods and tools used for analyzing malware that targets Linux-based IoT devices. It covers both static and dynamic analysis techniques and emphasizes the role of ML in classifying and detecting IoT malware. Additionally, it discusses various data collection methods for threat analysis, including the use of IoT honeypots, and explores the application of machine learning techniques to categorize and identify malware based on specific features. 
Similarly, Gaurav et al.~\cite{gaurav2023comprehensive} explores the use of machine learning techniques for detecting malware in IoT devices within enterprise systems. It comprehensively reviews various detection approaches, including static, dynamic, and hybrid methods. 
Notable limitations of these works are that (1) most of them lack an in-depth discussion of deep learning techniques for IoT malware analysis, and (2) they lack addressing malware across various XIoT domains, such as Android, IIoT, IoMT, and IoV.

Another surveys ~\cite{gopinath2023comprehensive, al2020survey} focus on deep learning-based methods to enhance security within the IoT domain, particularly for detecting malware across various platforms such as IoT, mobile, and Windows environments. \cite{gopinath2023comprehensive} traces the evolution of malware detection techniques from traditional methods to advanced deep learning approaches, providing an in-depth analysis of various models and techniques. The study also discusses different types of malware, including ransomware, Advanced Persistent Threats (APTs), and mobile malware.
Similarly, Al-Garadi et al.~\cite{al2020survey} explore the application of machine learning and deep learning techniques to enhance security in the IoT domain. Their work addresses a range of cyber and physical security threats to IoT, including those targeting physical components, networks, cloud infrastructures, web services, and applications, as well as emerging attack vectors. Additionally, the paper reviews how machine learning and deep learning methods can be employed to strengthen security measures against potential attacks.
However, both surveys have notable limitations: (1) insufficient coverage of deep learning-based methods for detecting malware in the IoMT and IoV domains, and (2) a lack of citations related to the use of deep learning techniques for malware detection.



Ngo et al.~\cite{ngo2020survey}  focuses on static feature analysis for detecting IoT malware. The paper reviews various static methods used in identifying malware across IoT devices such as OpCodes, file headers, and grayscale images and evaluates their effectiveness—besides a comparison of non-graph-based and graph-based detection methods. However, the paper has limitations in different places, such as static analysis, which might be less effective against sophisticated or newly emerging malware that can evade static detection.

Aslan et al.~\cite{aslan2020comprehensive} discuss various malware detection methods, including signature-based, heuristic, behavior-based, cloud-based, and deep learning approaches. They provide detailed explanations of the underlying technology, applications, advantages, and weaknesses of each approach. However, the paper has some limitations; notably, the flow chart of the malware detection process is confusing and not well-explained, and it improperly groups deep learning-based, signature-based, behavior-based, IoT-based, cloud-based, and mobile-based methods into a single category for feature extraction.


Raju et al.~\cite{raju2021survey} focuses on malware threat detection across various CPU architectures and operating systems. It presents a taxonomy of malware detection techniques, particularly emphasizing static and dynamic analysis methods tailored for multi-architecture environments. For static analysis, malware detection is categorized into metric-based, graph/tree-based, sequence-based, and interdependence approaches; for dynamic analysis, it is based on trace and usage features. However, the paper has limitations; notably, the paper does not cover all the taxonomy fields mentioned in the figure, and it does not discuss deep learning-based methods in XIoT domains.

Wazid et al.~\cite{wazid2019iomt} address the challenges posed by diverse CPU architectures and operating systems in IoT environments, complicating the implementation of threat detection and mitigation strategies. They discuss different malware types and the corresponding symptoms indicating their presence and also cover the architecture of the IoT communication environment. Although the paper title is supposed to focus more on the IoMT domain, it does not cover any IoMT research involving deep learning methods for malware analysis and does not discuss the potential challenges of implementing deep learning techniques, which require extensive data. Additionally, the taxonomy outlined in the paper, which includes key management, access/user control, intrusion detection, and user/device authentication, is generic and applicable to various XIoT domains, not specifically tailored to IoMT or IoT. The paper also lacks in-depth analysis specific to the IoMT domain.

\begin{figure}[!t]
    \centering
    \includegraphics[width=0.4\textwidth]{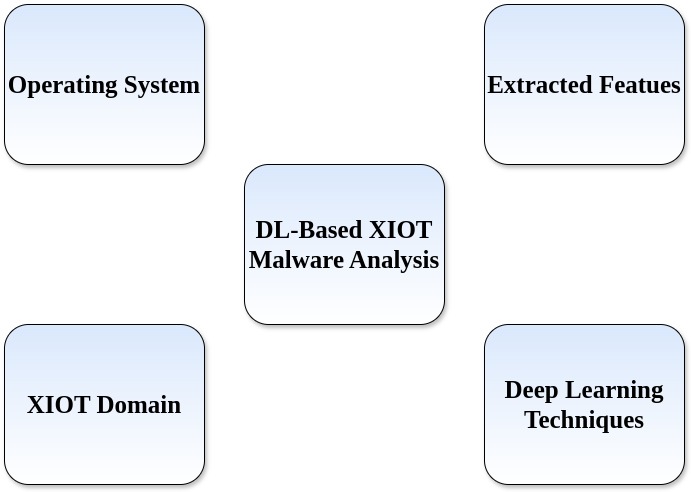}
    \caption{Overview of DL-Based XIoT Malware Analysis}
    \label{fig:Overview_XIoT}
\end{figure}

\subsection{\textbf{Taxonomy of DL-based XIoT Malware Analysis}}
\begin{figure*}[!t]
    \centering
    \includegraphics[width=0.7\textwidth]{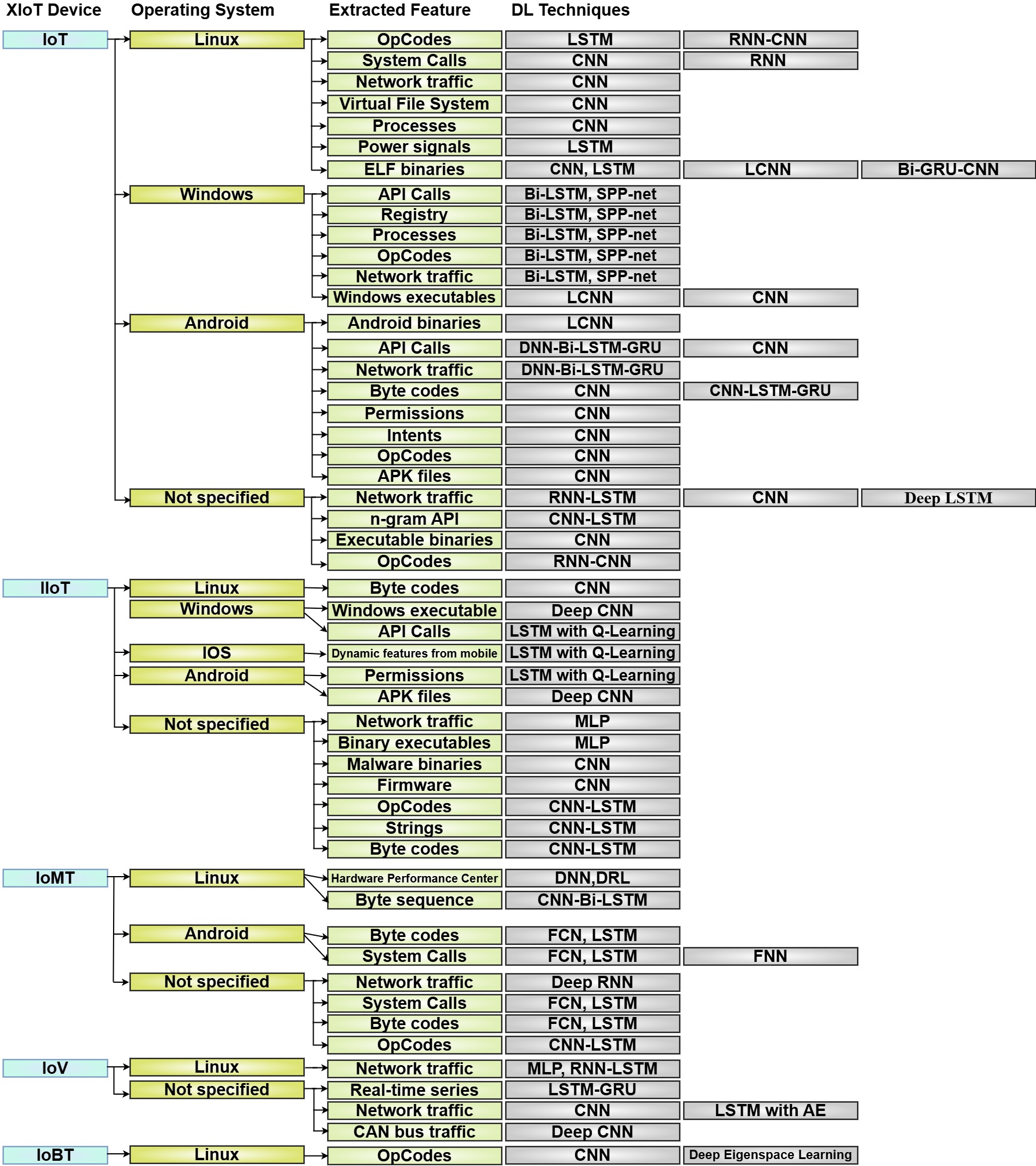}
    \caption{Taxonomy of the Research}
    \label{fig:Taxonomy}
\end{figure*}

Within the vast domain of the XIoT, malware poses a significant threat, demanding advanced big data analysis tools as well as effective mitigation strategies. Deep learning has emerged as a powerful tool for identifying these threats, thanks to its sophisticated pattern recognition capabilities. This subsection presents the taxonomy of DL methodologies adopted for malware analysis across various XIoT domains. Fig.\ref{fig:Overview_XIoT} illustrates the central component of deep learning-Based  XIoT Malware Analysis, influenced by four key elements. The "Operating System" block indicates the role of the domain in which the malware operates and is analyzed. "Extracted Features" refers to the specific characteristics or data points derived from malware samples, serving as critical inputs for deep learning models. The "Deep Learning Techniques" block emphasizes the use of advanced deep learning models to analyze and detect malware within the XIoT domain. Lastly, the "XIoT Domain" highlights the specific context, focusing on a wide range of connected devices. Fig.\ref{fig:Taxonomy} provides the taxonomy of our research, categorizing the XIoT domains into IoT, IIoT, IoMT, IoV, and IoBT. Each of these domains may operate under specific operating systems, which in turn influence the process of feature extraction. For instance, both Linux and Android operating systems might utilize similar feature extraction techniques, such as extracting Opcodes and Network Traffic. However, certain features are unique to specific operating systems; for example, the extraction of Permissions is specific to the Android operating system. Once the features are extracted, the choice of deep learning-based techniques for malware detection is made based on the nature of the extracted features and the corresponding XIoT domain. These techniques are tailored to enhance the accuracy and efficiency of malware detection across different XIoT domains.

\subsection{\textbf{Our Contribution}}
The main contributions of the paper are three-fold:
\begin{itemize}
    \item We focus research studies on malware analysis across various sectors of the XIoT, including IoT, IIoT, IoMT, IoBT, and IoV.
    \item We provide a taxonomy based on four critical attributes shared between all XIoT domains, namely domain, operating system, extracted features and DL techniques.
    \item We particularly focus on research studies that employ deep learning techniques for malware detection in XIoT domains, aiming to consolidate all related DL-based research within a single comprehensive survey.
\end{itemize}

\subsection{\textbf{Survey Methodology}}
We collected relevant studies on DL-based malware analysis on multiple XIoT domains from various databases, including Google Scholar, IEEE Xplore, ACM Digital Library, ScienceDirect, and SpringerLink. The keywords used in our search, identified by quotation marks, include ``malware detection'', ``malware classification'', ``malware analysis'', combined with keywords such as ``deep learning'', ``Internet of Things (IoT)'', ``Industrial Internet of Things (IIoT)'', ``Internet of Battlefield Things (IoBT)'', ``Internet of Medical Things (IoMT)'', and ``Internet of Vehicles (IoV)''. Our survey covers publications from the years 2017 to 2024.

\subsection{\textbf{Organization of the Survey}}
The rest of this survey is organized as follows.
Section~\ref{sec:background} introduces the fundamental concepts of the XIoT domain, which includes the IoT, IIoT, IoMT, IoV, IoBT, and Android platforms. Additionally, it discusses various deep learning techniques used in malware analysis, along with malware analysis approaches. 
Section~\ref{sec:xiot_malware_overview} discusses the factors that directly influence malware analysis in the XIoT domain, including ELF and PE file formats, CPU architecture, and operating systems. 
Section~\ref{sec:dl_xiot_malware_analysis} presents a comprehensive review of research studies that utilize deep learning techniques for analyzing malware within the XIoT domain.
Finally, Section~\ref{sec:conclusion} summarizes and concludes the work.

\begin{figure*}[!t]
    \centering
    \includegraphics[width=\textwidth]{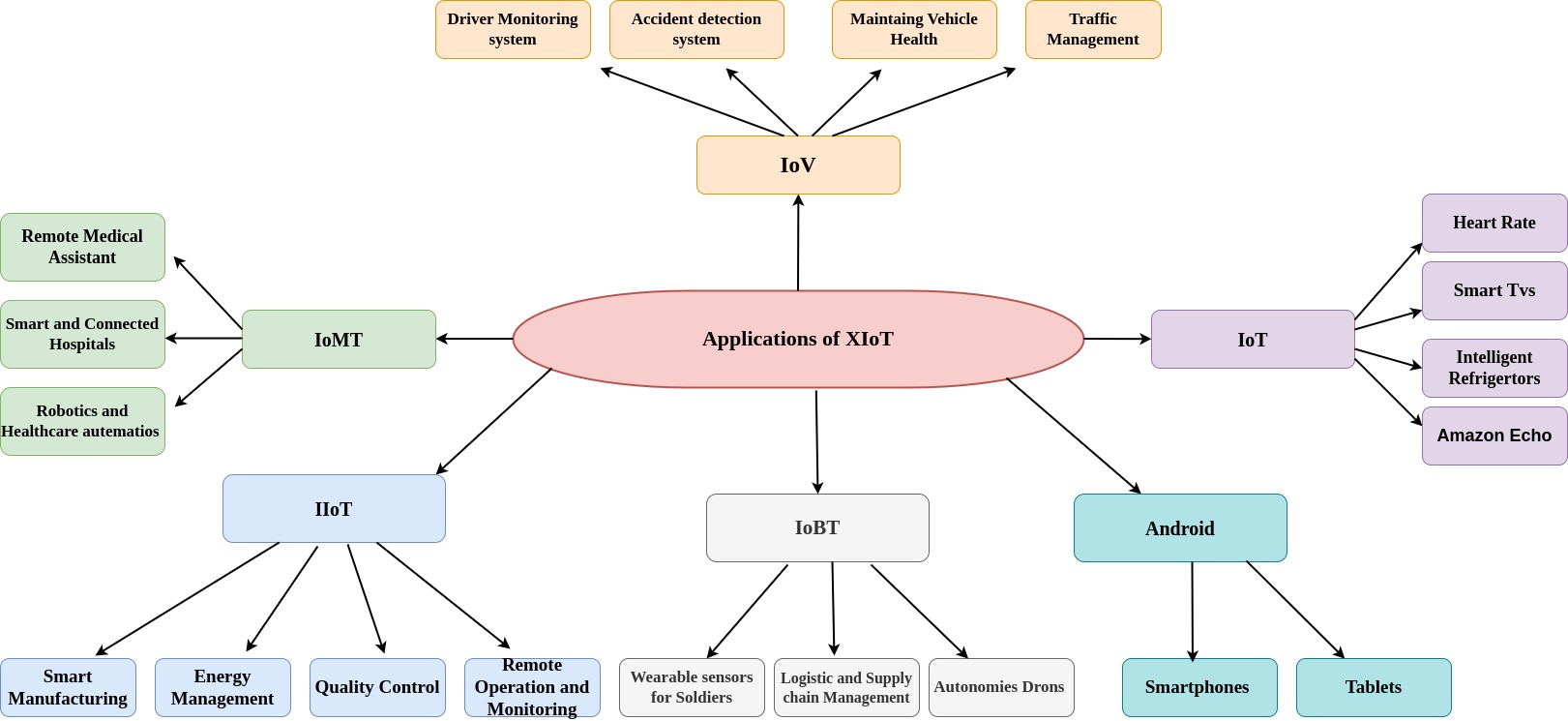}
    \caption{Applications of XIoT}
    \label{fig:XIoT}
\end{figure*}

\section{\textbf{Background}}
\label{sec:background}
This section provides a background on three key areas: the XIoT, deep learning techniques, and malware analysis approaches. 
\subsection{\textbf{Extended Internet of Things (XIoT)}}
The various applications of XIoT span across multiple domains, including IoT, IIoT, IoMT, IoV, and IoBT, as illustrated in Fig.~\ref{fig:XIoT}. These applications include diverse sectors, from healthcare and industrial automation to vehicular technology and battlefield operations.

\subsubsection{Internet of Things (IoT)}
IoT is an expansive network of interconnected devices that enhance everyday life by seamlessly integrating digital intelligence into the physical world. From home automation systems that allow for remote control of household appliances—improving energy efficiency and security—to wearable technologies that monitor health metrics like heart rate and activity levels to provide insights into personal wellness, IoT is revolutionizing how we interact with our environment. In consumer electronics, IoT-enabled devices such as smart TVs and intelligent refrigerators not only offer unprecedented convenience but also adapt to user behaviors to optimize functionality and enhance user experience. These IoT systems communicate through the Internet, exchanging data to automate, optimize, and make more intelligent decisions, which significantly contributes to a more connected and efficient world.

\subsubsection{Industrial Internet of Things (IIoT)} 
IoT technologies in industrial settings, known as the IIoT, have significantly altered the architecture of Industrial Automation and Control Systems (IACS). Historically, IACS were largely isolated from mainstream digital networks, utilizing firewalls and demilitarized zones for security when connectivity was necessary \cite{boyes2018industrial}. However, the emergence of IIoT has led to increased connectivity and inherent architectural changes within these systems, facilitating a closer interaction between industrial systems and cyber-physical systems under the broader umbrella of Industry 4.0 \cite{boyes2018industrial}. IIoT's role in smart manufacturing, energy management, quality control, and remote operation is crucial for advancing industrial efficiency and safety.

\subsubsection{Internet of Medical Things (IoMT)}
IoMT is an innovative development that merges telehealth technologies with IoT, significantly enhancing medical care efficiency and patient comfort. IoMT incorporates advanced IoT frameworks with medical devices to facilitate healthcare delivery through faster, cost-effective, and more personalized solutions. An elaborate architecture supports IoMT operations, consisting of cloud, fog, and end-user layers that support real-time health data processing and analysis. This structure harnesses emerging technologies such as Physically Unclonable Functions (PUF), Blockchain, Artificial Intelligence, and Software-Defined Networking to enhance security, privacy, and performance efficiency \cite{razdan2022internet}. IoMT applications like remote medical assistance, smart and connected hospitals, and robotics for healthcare are vital components of the XIoT landscape.

\subsubsection{Internet of Battlefield Things (IoBT)}
IoBT represents a pivotal evolution in cyber-physical systems, aiming to enhance battlefield resilience through advanced connectivity and automation. Central to the IoBT is its ability to resist and adapt to a diverse array of threats, significantly more varied than those faced by its civilian counterparts. This robustness is achieved by integrating heterogeneous devices — including sensors, actuators, and intelligent algorithms — all designed to function under the extreme conditions of warfare \cite{abdelzaher2018toward}. Wearable sensors for soldiers, logistics, and autonomous drones are among the critical applications of IoBT in enhancing battlefield capabilities.

\subsubsection{Internet of Vehicles (IoV)}
IoV focuses on enhancing vehicular communication and integrates various data-driven services to support safer and more efficient roadways \cite{karim2022architecture}. However, IoV is characterized by a dynamic and automated interplay between vehicles, infrastructure, and users, facilitated through sophisticated sensory and networking technologies. IoV's pivotal role in modern transportation is the ability to allow vehicles to interact with their environment and with each other in real-time. This interaction is supported by a multi-layered architecture comprising sensory, network, and application layers, each playing a crucial role in data collection, processing, and application, respectively. Such comprehensive connectivity is aimed at improving traffic management, minimizing the risks of accidents, and optimizing logistics and vehicle maintenance through predictive analytics.

\subsection{\textbf{Deep Learning Techniques}}
\subsubsection{Multi-Layer Perceptron}
The Multi-Layer Perceptron (MLP) or Deep Neural Network (DNN) is a fundamental type of feedforward artificial neural network crucial to the architecture of deep learning systems. As a supervised learning model, the MLP comprises multiple layers including an input layer, one or more hidden layers, and an output layer, forming a fully connected network. This connectivity means that each neuron in a given layer links directly to every neuron in the following layer. The primary role of the input layer is to receive and normalize input data, while the hidden layers focus on processing these inputs through a series of computational steps. The output layer is responsible for making predictions or decisions based on the aggregated and processed information from the hidden layers \cite{shiri2023comprehensive}.

\subsubsection{Convolutional Neural Networks}
Convolutional Neural Networks (CNNs) are a sophisticated class of feedforward neural networks that have revolutionized various fields by enabling automatic feature extraction without manual intervention, a significant advance over traditional methods. Primarily inspired by the human visual perception system, CNNs are particularly powerful in image classification, object detection, speech recognition, and even bioinformatics and time series prediction. The architecture of a CNN typically involves several layers, including convolutional layers, pooling layers, and fully connected layers. In the convolutional layer, learnable filters extract different levels of features from the input, for instance, simple textures and patterns in earlier layers and more complex details in deeper layers. Following this, the pooling layer helps reduce dimensionality and computational load while maintaining the robustness of the features extracted. This layer also aids in making the detection process stable to minor changes in the position and orientation of recognized objects. The CNN usually ends with a fully connected layer that acts like a classifier, integrating learned features into final predictions \cite{shiri2023comprehensive}.

\subsubsection{Spatial Pyramid Pooling Networks}
Spatial Pyramid Pooling Networks (SPP-net) introduce the spatial pyramid pooling layer into a CNN architecture to remove the requirement for a fixed-size input image. This layer performs a pooling operation that can accept inputs of any size and produce outputs of a fixed size so that they can be fed into the fully connected layers downstream without needing to resize or crop the input images first. By employing multi-level pooling from the entire image, the network retains the ability to recognize objects across various scales and deformations. SPP allows the network to be more versatile in handling images of different dimensions, enhancing its usability in practical applications where input sizes vary significantly. The SPP layer is implemented by dividing the feature map that comes from the last convolutional layer into spatial bins at multiple levels of resolution, which correspond to the layers of the pyramid. These bins are then pooled (typically using max pooling) to generate a fixed-length output vector for each level of the pyramid, which is concatenated to form the final output of the SPP layer. \cite{he2015spatial}.

\subsubsection{Recurrent Neural Networks}
Recurrent Neural Networks (RNNs) are a sophisticated class of deep learning models designed to process sequential data using internal memory to capture temporal dependencies. Unlike conventional neural networks, which perceive each input as independent, RNNs can consider the order of inputs, which is essential for tasks involving sequential information such as natural language processing, speech recognition, and video classification. The core feature of RNNs is the loop mechanism that allows them to perform the same operation on every sequence element, with the output of each step being dependent on the previous outputs. This is achieved through a memory state or hidden layer, which integrates the current input with the previous state using an activation function such as the hyperbolic tangent. Although traditional RNNs are limited by their short-term memory, leading to challenges in learning from long input sequences \cite{shiri2023comprehensive}.

\subsubsection{Long Short-Term Memory}
Long Short-Term Memory (LSTM) networks, a specialized form of RNNs, were introduced to overcome the challenge of capturing long-term dependencies in sequence data, a limitation faced by traditional RNNs. LSTMs have significantly shaped the landscape of deep learning through their unique architecture, designed to enhance memory retention over extended periods. Central to the LSTM model are three distinct gates: the input gate, the forget gate, and the output gate, which collaboratively manage the flow of information within the unit. These gates control the updating of the internal state, the extent to which previous states are retained or discarded, and the influence of the internal state on the output. This configuration allows LSTMs to maintain a durable link to past information, facilitating more effective learning and prediction in tasks involving long sequence data. Moreover, the architecture typically includes a series of these units through which both the current input and the output from the previous timestep are processed sequentially, ensuring dynamic updating of states throughout the network \cite{shiri2023comprehensive}.

\subsubsection{Bidirectional Long Short-Term Memory}
Bidirectional Long Short-Term Memory (Bi-LSTM) enhances the traditional LSTM architecture that enhances sequence modeling by considering past and future contexts. Standard LSTMs process inputs in a forward direction only; Bi-LSTMs address this limitation by incorporating a backward direction, allowing for a comprehensive understanding of temporal dependencies. Specifically, a Bi-LSTM consists of two parallel LSTM layers: one processes the input sequence from left to right (forward LSTM layer) and the other from right to left (backward LSTM layer). This setup enables the model to gather information from both directions simultaneously. During the training phase, the forward and backward LSTM layers independently extract features and update their internal states based on the sequence inputs. Each layer outputs a prediction score at every time step, which is then combined using a weighted sum to produce the final result \cite{shiri2023comprehensive}.

\subsubsection{Gated Recurrent Unit}
The Gated Recurrent Unit (GRU) is another variant of the RNN architecture designed to address the short-term memory issue while offering a simpler structure compared to the LSTM. By combining the input and forget gates into a single update gate, the GRU provides a more streamlined design without a separate cell state. It consists of three main components: the update gate, the reset gate, and the current memory content. These components enable the GRU to selectively update and utilize information from previous time steps, thus capturing long-term dependencies effectively. The update gate controls how much past information is retained and combined with current input, while the reset gate determines how much past information should be forgotten. The current memory content, influenced by the reset gate, contributes to the candidate activation for updating the final memory state. Although GRU simplifies the number of tensor operations, thus potentially allowing for faster training, the choice between GRU and LSTM depends on the specific use case and the nature of the task at hand \cite{shiri2023comprehensive}.

\subsubsection{Generative Adversarial Networks}
Generative Adversarial Networks (GANs) are a powerful class of neural network architectures designed for generative modeling. These networks facilitate the generation of realistic and novel samples by leveraging a dual-model structure that includes a generative and discriminative models. The generative model produces images that mimic real-life examples, whereas the discriminative model evaluates these images to distinguish between genuine and synthetic outputs. This setup enables a dynamic, adversarial interaction where the generator and discriminator iteratively improve through a competitive game-like scenario. In practice, GANs first employ the generator to create content that emulates the real data distribution, then train the discriminator better to identify the differences between real and generated images. This training process involves backpropagation and repeated presentation of datasets to enhance the discriminator's accuracy and the generator's ability to deceive. Typically, the discriminator is constituted by CNNs, while deconvolutional networks are used for the generator, refining their capability to produce increasingly sophisticated and indistinguishable images from real ones \cite{shiri2023comprehensive}.

\subsection{\textbf{Malware Analysis Approaches}}
In this subsection, we explore the four fundamental techniques for malware analysis: static, dynamic, hybrid, and online approaches. These methods range from analyzing malware code without execution to monitoring its behavior in real-time environments.

\subsubsection{\textbf{Static Analysis}}
static malware detection involves using techniques that analyze the characteristics and behavior of malware without executing the code. There are different types of static detection methods, including CFG analysis, opcode-based techniques, and analysis of byte sequences from executable files \cite{bobrovnikova2022technique} \cite{darabian2020opcode} \cite{wan2020efficient}. However, the use of static analysis in IoT malware detection is significant due to IoT devices' limited resources and processing power, as well as the diversity of device types and operating systems \cite{ali2023machine}. Furthermore, static analysis is crucial for understanding the behavior of IoT malware without adding more effort to run the code. Moreover, the static analysis techniques in IoT malware detection can be implemented through machine learning and deep learning techniques, as they provide valuable information about the behavior and characteristics of malware without executing the code. Integrating modern ML and DL techniques with static analysis methods provides practical solutions to the new IoT malware variants \cite{eboya2020intelligent}.

\subsubsection{\textbf{Dynamic Analysis}}
The dynamic malware detection approach is characterized by identifying malware in IoT devices through the execution of malware samples and monitoring their behavior. This approach is particularly essential for detecting new and unrecognized malware types, which traditional signature-based detection methods often fail to recognize \cite{lim2014mal, aslan2020comprehensive}. Conducted within a controlled, isolated environment like a sandbox, dynamic analysis allows for the comprehensive extraction of key behavioral features such as file interactions, process behaviors, and network communications \cite{ki2015novel}. Furthermore, machine learning and deep learning techniques play an important role by facilitating the extraction and analysis of features during code execution, which provides a deeper understanding of the malware's true behavior \cite{wang2021novel}. These methods have been tested using both actual malware samples and benign programs within a controlled environment to monitor and assess malware activities \cite{li2018evaluation}. Additionally, the behavioral patterns derived from dynamic analysis are essential in detecting and categorizing unknown malware into known groups \cite{talukder2020tools}.

\subsubsection{\textbf{Hybrid Analysis}}
The hybrid analysis approach combines static and dynamic analysis to enhance detection accuracy. This method leverages the strengths of both: analyzing malware code without execution (static analysis) and observing malware behavior during execution in a controlled environment (dynamic analysis). For instance, research \cite{tong2017hybrid} has demonstrated the efficiency of this approach by collecting execution data from malware and benign applications to distinguish patterns related to system calls for file and network access. By doing so, they establish distinct sets of malicious and normal behaviors that differentiate these patterns. Afterward, when examining new applications, dynamic analysis is utilized to gather system call data, which is then compared with these predefined pattern sets to determine the application's nature. Similarly, \cite{gaba2022analysis} integrates static and dynamic analyses; the static part involves reverse engineering to understand the malware structure and potential impact, while the dynamic part observes its behavior upon execution.

\subsubsection{\textbf{Online Analysis}}
The online malware detection approach is characterized by continuous system monitoring to preemptively identify and mitigate malware. This approach is different from both static and dynamic analysis approaches. Specifically, static analysis examines executable code to create malware signatures, while dynamic analysis involves the execution of programs within controlled settings to monitor their behavior. In online malware detection, despite preventative strategies, malware may penetrate systems, which requires continuous attention \cite{kimmel2021recurrent}. However, various advanced techniques focus on online malware detection, including deep learning and ensemble methods. However, these methods typically merge elements from both static and dynamic analysis to develop efficient and scalable models capable of accurately identifying malware \cite{yuan2016droiddetector} \cite{yan2018detecting} \cite{ota2017deep}. Furthermore, online malware detection employs techniques such as virtual machines, dynamic binary instrumentation, information flow tracking, and software anomaly detection to continuously monitor and identify potential security threats \cite{ozsoy2015malware}. Other efforts in online malware detection \cite{abdelsalam2019online} \cite{mcdole2020analyzing} incorporate performance metrics and process-level data, alongside behavioral analyses including examining CPU, memory, and disk usage to determine the existence of malware.

\section{\textbf{Overview of XIoT Malware}}
\label{sec:xiot_malware_overview}
This section provides an overview of malware within domain of the XIoT, covering widely used file format to various CPU architectures and operating systems integral to XIoT devices. Additionally, we present the main features that are essential for effective malware analysis.

\begin{figure}[!t]
\centering
\includegraphics[width=1\linewidth]{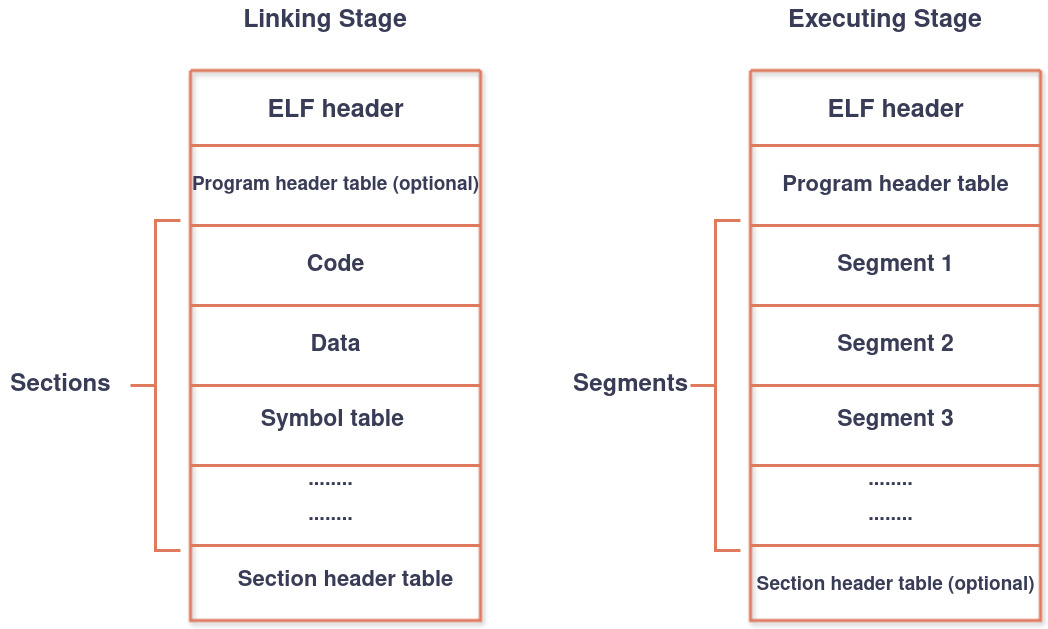}
\caption{Executable and Linkable Format (ELF)}
\label{fig:elf}
\end{figure}

\subsection{\textbf{Executable File Formats}}
\subsubsection{\textbf{ELF File Format}}
The Executable and Linkable Format (ELF) is a standard file format that is notable for its flexibility in supporting a wide range of CPU architectures, including x86, ARM, and RISC-V \cite{r1} \cite{r2}. ELF files incorporate various types, such as executable files, relocatable files, shared object files, core dump files, and specific processor supplement files. Each type contains crucial data, including the program header table, which contains entries describing the segments used for loading the file into memory, and the section header table, which defines the sections within the file, such as code, data, and symbol tables \cite{kell2016missing}.

As shown in Fig.~\ref{fig:elf}, the ELF format is divided into two main stages: the Linking Stage and the Executing Stage. During the linking stage, the ELF file is organized into sections, each serving a specific purpose. These sections include the Code (executable instructions), Data (initialized variables), Symbol Table (linking references such as function names and variable locations), and Section Header Table (which lists all sections within the file). This structure ensures that the linker can correctly combine multiple object files into a single executable or shared object. In contrast, the executing stage focuses on Segments, which are derived from the sections during the linking process and are loaded into memory when the program runs. These segments typically include the executable code, data, and other essential runtime information. The ELF header and program header table play a critical role in mapping these segments into memory, ensuring that the operating system can correctly execute the program.

In the XIoT malware domain, the ELF file format is essential due to the widespread use of Unix-based systems in IoT devices, which primarily use ELF as their standard format for binary files \cite{sivakumaran2021argxtract}, making ELF the default executable file format. Additionally, the importance of ELF files is highlighted in the dynamic analysis of IoT malware, since behaviors linked to memory, network, virtual file systems, processes, and system calls are derived from ELF files for effective malware detection \cite{jeong2022massive}. The classification of IoT malware often involves techniques such as analyzing OpCode sequences extracted from disassembled ELF files or evaluating features in the Control Flow Graph (CFG) generated from these sequences.

\subsubsection{\textbf{PE File Format}}
\begin{figure}[!t]
\centering
\includegraphics[width=.55\linewidth]{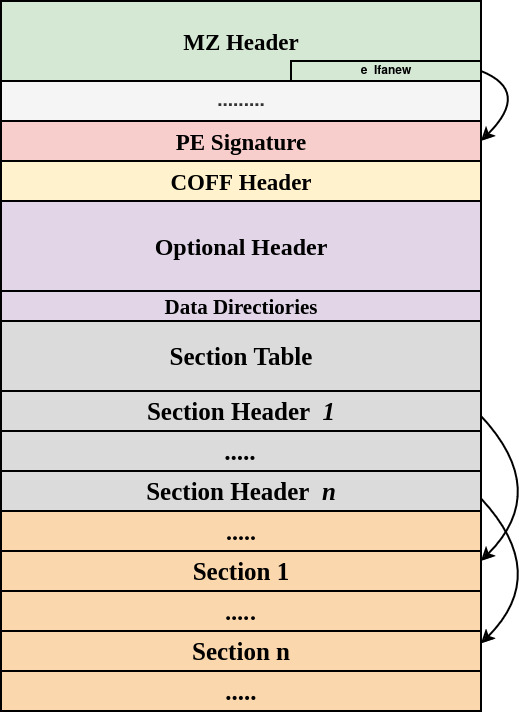}
\caption{Portable Executable File (PE)}
\label{fig:PE}
\end{figure}

The Portable Executable (PE) file format, as illustrated in Fig.~\ref{fig:PE}, is essential for executables, object code, DLLs, FON Font files, and other types used in 32-bit and 64-bit versions of Windows operating systems \cite{nisi2021lost}. This format includes a data structure crucial for the Windows OS loader to manage the wrapped executable code. It includes several key components: the MZ header, which ensures compatibility with MS-DOS systems and determines the offset for the start of the COFF Header; the COFF Header, which contains information about the architecture the executable is designed to run on, including support for a randomized base address; the Optional Header, which follows the COFF Header and provides detailed information such as the preferred base address of the executable and memory allocation specifics; the Data Directories within the Optional Header, detailing specific data structures used by the executable like import and export tables; and the Section Table, which defines various sections of the executable, such as code and data sections, crucial for the organization and execution of the binary. PE files are significant in malware analysis because they encapsulate executable code and metadata necessary for the Windows loader. Malware can be detected by analyzing the dynamic behavior of these files \cite{nguyen2024using}. Furthermore, analyzing static features extracted from the PE file headers, which include metadata about the executable code, such as machine architecture, code size, and system resources, provides a solid foundation for effective malware detection strategies \cite{lad2022improved}.

\subsection{\textbf{XIoT CPU Architectures}}
The selection of CPU architectures in XIoT devices, which include IoT as a sub-domain, is crucial, impacting various aspects such as resource allocation, task offloading, and botnet detection. XIoT devices often incorporate open-source code and employ a range of CPU architectures, including MIPS, MIPSEL, PPC, SPARC, ARM, MIPS64, sh4, and X86. Understanding how different CPU architectures influence XIoT systems is important because XIoT botnet samples are designed to operate across multiple CPU architectures, with some explicitly targeting specific types \cite{le2020v}.

The commercial market continuously upgrades XIoT architectures, including IoT, to reduce data transmission costs, latency, and bandwidth usage for diverse application requirements \cite{krishnamoorthy2021systematic}. Moreover, the security challenges posed by the diversity of CPU architectures in XIoT devices cannot be ignored, as vulnerabilities in these architectures can be exploited by malware, making XIoT devices, including IoT, susceptible to significant security issues \cite{wan2020iotargos}. However, choosing a lightweight CPU architecture is essential for optimizing XIoT device performance, resource utilization, and energy efficiency. Lightweight architectures, such as the ARM Cortex-M series, are specifically designed to meet the low-power and resource-constrained requirements of XIoT applications. These architectures balance performance and power consumption, making them ideal for XIoT deployments, including IoT-specific use cases \cite{xhonneux2023sub}.

\subsection{\textbf{XIoT Operating System Platforms}}
\begin{figure*}[!t]
    \centering
    \includegraphics[width=\textwidth]{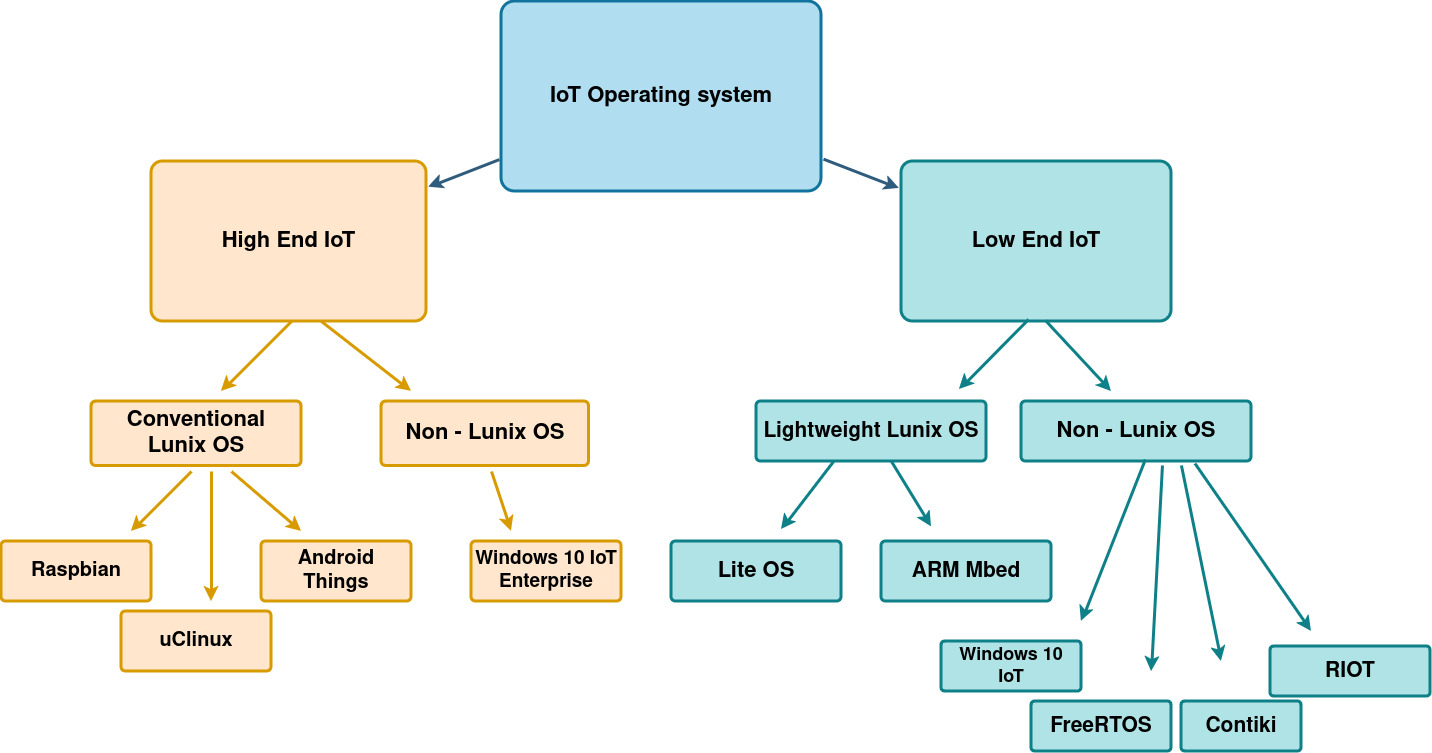}
    \caption{Different OS platforms applied on different IoT devices}
    \label{fig:IoT_OS}
\end{figure*}

The selection of an operating system for IoT devices is based on multiple considerations, including the specific use cases of the device, its resource constraints, security requirements, and the imperative for energy efficiency \cite{baddeley2018evolving}. As illustrated in Fig.~\ref{fig:IoT_OS}, devices at the higher end of the IoT spectrum may employ conventional operating systems such as Linux \cite{kumar2018clustering}, whereas devices at the lower end, characterized by stringent resource limitations, are often equipped with operating systems explicitly engineered for embedded IoT applications, including RIOT, Contiki, FreeRTOS, and ARM Mbed \cite{eschweiler2016discovre}. These lightweight operating systems are designed to address resource constraints and ensure real-time operation effectively. However, the operating system selection must meet the security paradigm within IoT frameworks, ensuring that the chosen operating system does not degrade the security integrity of IoT platforms and minimizes energy consumption while maximizing performance \cite{spillner2020rule}.

\subsection{\textbf{Data Representations for Malware Analysis}}
In this subsection, we discuss the feature extraction techniques that are used for malware detection within the XIoT domain. Table~\ref{analysis-techniques} presents data representations for various malware analysis techniques.

\subsubsection{Operation Code}
An Operation Code (OpCode) is a part of a machine language instruction that specifies the operation to be performed  \cite{cavanagh2013x86}. The OpCode is followed by one or more operands, which can be values, memory addresses, or processor registers, on which the operation should be executed. In the assembly language, OpCodes (e.g., "ADD", "SUB", or "MOV") play a crucial role, with each OpCode corresponding to a specific operation, such as adding two numbers, moving data from one location to another, or comparing two values. 

The OpCode sequences are utilized in malware detection methods, underscoring their importance in static analysis. Several research studies employ OpCodes in various ways to detect IoT malware. Some approches \cite{lee2023robust, moon2022evolved} categorize OpCode sequences based on functionality to distinguish IoT malware across different families for generating features that are applied to various machine learning models. Other \cite{lin2023hash} addresses the issue of identical OpCode sequences within function call graphs (FCGs) by employing hash functions to efficiently merge nodes with identical OpCodes, thereby enhancing the compactness and processing efficiency of the FCG representation. In addition, a different approach \cite{toan2022static} estimates the frequency of occurrences of OpCodes in OpCode sequences by applying the Term Frequency-Inverse Document Frequency (TF-IDF) model, which determines the significance of an OpCode based on its weight.

\subsubsection{Strings}

Strings are sequences of characters, texts, or commands within malware executables that can be analyzed to identify malicious behavior \cite{torabi2021strings}. In the IoT malware domain, strings are utilized in various research works. One approach \cite{torabi2021strings} characterizes IoT malware by identifying obfuscated samples and extracting meaningful strings, including commands and adversarial IP addresses. This technique is essential in revealing groups of IoT malware sharing same characteristics which helps in classifying malware families. Another approach \cite{lee2020cross} captures the typical characteristics of malware samples across different CPU architectures. The process involves extracting and encoding printable strings into numerical vectors, demonstrating discriminative capability in malware family classification.
Additionally, \cite{jeon2024static} converts strings into a visual format through vectorization; Mal3S applies a SPP-net model to analyze feature images derived from strings. Furthermore, \cite{wang2021malware} transforms behavioral characteristics into a map of sequential word vectors, applying a CNN to identify malicious software. This underscores the value of using strings as word embeddings to represent program behaviors, highlighting the potential of strings in malware detection within cloud computing infrastructures.

\subsubsection{ELF File header}
The importance of static analysis, particularly in analyzing the ELF file header, is highlighted in several research studies. A study \cite{ravi2022static} extracts static features from ELF executables and employs chi-square feature selection to accommodate the constrained resources of IoT devices. It demonstrates the ability to identify malware while maintaining minimal computational requirements. Another framework integrates static analysis of ELF headers and OpCode sequences \cite{tien2020machine}, leveraging machine learning techniques for improved detection and classification of malware. This method involves extracting features from ELF headers to gain a deep understanding of the binary's architecture and behavior. When combined with OpCode sequences, this approach enables a strong distinction between malicious and benign software. Another approach \cite{wan2020iot} explores the extraction of byte sequences from ELF file entry points, a technique that leverages the structured format of ELF files for malware identification. This method emphasizes the importance of ELF header information in distinguishing malware from benign software across a variety of CPU architectures.

\begin{table}[t]
\centering
\caption{Data representation used for malware analysis}
\label{analysis-techniques}
\begin{tabular}{@{}lcccc@{}}
\toprule
\multirow{2}{*}{\textbf{Data Representation}} & \multicolumn{4}{c}{\textbf{Analysis Techniques}}                   \\
                                              & \textbf{Static} & \textbf{Dynamic} & \textbf{Hybrid} & \textbf{Online} \\ \midrule
ELF Header                                    & \cmark          & \xmark           & \cmark          & \xmark          \\
Strings                                       & \cmark          & \cmark           & \cmark          & \xmark          \\
Opcode                                        & \cmark          & \cmark           & \cmark          & \xmark          \\
Control Flow Graph                      & \cmark          & \cmark           & \cmark          & \xmark          \\
Byte Sequence                                 & \cmark          & \xmark           & \cmark          & \xmark          \\
API Calls                                     & \cmark          & \cmark           & \cmark          & \cmark          \\
Memory Usage                                  & \xmark          & \cmark           & \cmark          & \cmark          \\
Network Traffic                               & \xmark          & \cmark           & \cmark          & \cmark          \\
Instruction Trace                             & \xmark          & \cmark           & \cmark          & \xmark          \\
System Processes                              & \xmark          & \cmark           & \cmark          & \cmark          \\ \bottomrule
\end{tabular}
\end{table}

\subsubsection{Control Flow Graphs}
Control Flow Graphs (CFGs) are widely utilized in code analysis and optimization due to their clear and intuitive representation of control dependencies within a program \cite{bai2009detecting}. One method transforms each malware sample into a graph representation \cite{alasmary2019analyzing}, facilitating the extraction of static features that are representative of IoT binary files. These features include graph-related attributes like nodes and edges, which encapsulate the execution flow of a program. Specifically, the CFG captures the connections between different code segments, where nodes represent the program's basic blocks or decision points, and the edges illustrate the flow from one block to another based on the execution conditions. By analyzing these features using machine learning models, the system can effectively differentiate between IoT files and Android files, as well as distinguish IoT malware from benign IoT files. Another approach, CDGDroid \cite{xu2018cdgdroid}, combines CFGs with Data Flow Graphs (DFGs) to create detailed semantic representations of Android applications. These graphs are encoded into matrices and utilized as features for training a CNN to detect malware. The horizontal combination of CFG and DFG features has been found to deliver superior performance in classification accuracy, outperforming other methods in detecting Android malware across various datasets. Another study \cite{gao2022malware} extracts these graphs from PE files to identify malware. The CFGs are processed using the MiniLM pre-trained language model to generate node features, which are then transformed into a compact vector using a Graph Isomorphism Network (GIN) and classified via a Multilayer Perceptron (MLP) to differentiate between malicious and benign software.

\subsubsection{API calls}
API calls analysis is increasingly recognized as a method for detecting malware in IoT environments. The Mal3S approach \cite{jeon2024static} utilizes static analysis to analyze API calls and transforms malware indicators into analyzable images. The MalDozer framework \cite{karbab2018maldozer} highlights the effectiveness of deep learning by differentiating between benign and malicious app behaviors by extracting API call sequences from the app's DEX files. EveDroid \cite{lei2019evedroid} focuses on behavioral analysis by mapping API calls to event groups, thus facilitating a dynamic analysis of malware actions and using neural networks to achieve precise malware detection on IoT devices by capturing the semantic relationships between events. API-MalDetect \cite{maniriho2023api} integrates advanced natural language processing (NLP) techniques with hybrid neural networks to analyze Windows malware through identifying patterns within API call sequences.

\subsubsection{Memory Usage}
Memory usage patterns play a crucial role in the dynamic analysis of malware within nested cloud environments \cite{jeon2020dynamic}. The approach discussed in this research study \cite{jeon2020dynamic} highlights the importance of analyzing memory usage patterns as part of a dynamic analysis framework. By analyzing memory behaviors alongside other key system behaviors such as network communications, virtual file system activities, process executions, and system calls, the research captures a holistic view of an IoT device's operational state. This multifaceted behavioral data is then transformed into a visual format and processed through a CNN model for IoT malware classification.

\subsubsection{Network Traffic}
Network traffic analysis has been identified as a critical technique for detecting IoT malware, and it has been highlighted by various methodologies in recent research works. One study \cite{wei2023comparing} applies Random Forest and Naïve Bayes machine learning algorithms to differentiate malware from benign traffic, emphasizing the importance of selecting key network traffic features. Another study \cite{nimalasingam2022detection} proposes a model based on the forensic analysis of network traffic features, highlighting the vital role of detailed network traffic characteristics in identifying IoT malware. Moreover, a multitask deep learning approach \cite{ali2022effective}, leveraging LSTM networks for the detection and classification of IoT malware through behavioral traffic analysis.

\subsubsection{Instruction Trace}
The trace instruction can be utilized in various ways, such as monitoring network traffic, OpCode instruction monitoring, and process tracking. Research work \cite{al2023iot} integrates deep learning and NLP techniques to analyze OpCode instructions representing various CPU architecture attacks. Similarly, research \cite{hu2020exploit} analyzes IoT malware by utilizing disassembled instruction sequences as semantic features, building a hierarchical model to retain the original instruction hierarchy, and using deep learning techniques for feature extraction and classification. Additionally, a method detailed in \cite{bai2022real} verifies IoT device instructions in real-time through side-channel analysis, employing the RASC system—a small, low-cost external monitor that analyzes power consumption patterns to detect malicious alterations at the instruction level. This approach enables non-intrusive, fine-grained monitoring and real-time malware detection by comparing power traces against established models of benign behavior.

\subsubsection{System Processes}
In the domain of malware detection, several research studies focus on system processes. One research study \cite{abdelsalam2019online} introduces an approach to online malware detection in cloud environments by utilizing a shallow CNN in conjunction with auto-scaling systems. This method leverages the dynamic nature of auto-scaling in the cloud, which adjusts the number of active virtual machines based on demand. By deploying a malware detection system that operates in real-time, the system continuously monitors for anomalous behaviors across these virtual machines. The approach leverages process-level performance metrics to train the CNN, enabling the model to detect unusual patterns potentially indicative of malware. Another study \cite{jeon2020dynamic} focuses on dynamically analyzing malware-related activities, such as memory utilization, network behavior, process executions, and system calls. The data from these processes is transformed into images that are then classified and trained using a CNN, specifically ZFNet, to detect known and emerging IoT malware variants effectively. Furthermore, a different approach \cite{rhode2018early} centers around the initial system activities immediately upon file execution. By capturing snapshots of CPU usage, memory usage, the number of processes running, and the maximum process ID assigned, this method employs an RNN to analyze these early-stage indicators to predict whether an executable is benign or malicious accurately.

\section{\textbf{DL-Based XIoT Malware Analysis}}
\label{sec:dl_xiot_malware_analysis}
This section discusses the studies we have reviewed and highlights the significant results associated with their approaches. In addition, the availability of datasets used by various works are extremely important and can facilitate the progress of work in this area for interested researchers. As such, we compiled a list of datasets used by different works in Table~\ref{tab:dataset}.
It summarizes the XIoT datasets used in DL-based XIoT malware analysis research works. 

\subsection{\textbf{Internet of Things (IoT)}}
Many research works utilize DL-based techniques in the IoT domain for malware analysis. Table~\ref{tab:iot} summarizes the research done on DL-based IoT malware analysis.
\begin{table*}[!ht]
    \centering
    \def\arraystretch{1.5}
    \caption{XIoT datasets and their sources used by research works. The dataset type M and B indicates malicious and benign, respectively. Note that some works fetched their data form public repositories such as VirusTotal and VirusShare, but the specific data samples used are not mentioned.}
    \rowcolors{2}{gray!25}{white}
    \begin{tabular}{|p{1.2cm}|p{3.8cm}|p{5.5cm}|p{0.8cm}|p{3cm}|}
    \rowcolor{gray!25}
    \hline
    \textbf{XIoT Domain} & \textbf{Dataset / Dataset Source} & \textbf{Dataset Link} & \textbf{Type} & \textbf{References} \\
    \hline
    
    IoT & Linux packages & \href{https://pkgs.org/}{pkgs.org} & B & \cite{haddadpajouh2018deep} \\ \hline
    
    IoT, IoMT & VirusTotal & \href{https://www.virustotal.com/}{virustotal.com} & M & \cite{dib2021multi}, \cite{chaganti2022deep}, \cite{haddadpajouh2018deep} , \cite{he2023image} \\ \hline

    IoT, IIoT, IoMT & VirusShare & \href{https://virusshare.com/}{virusshare.com} & M & \cite{dib2021multi}, \cite{chaganti2022deep}, \cite{vasan2020mthael}, \cite{sharma2023windows},\cite{ding2020deeppower}, \cite{ren2020end}, \cite{khowaja2021q}, \cite{amin2022deep}, \cite{he2023image}  \\ \hline
    
    IoT & IoTPoT & \href{https://sec.ynu.codes/iot/available_datasets/}{Available Dataset | YNU IoT} & M & \cite{shobana2020novel}, \cite{dib2021multi}, \cite{chaganti2022deep}, \cite{vasan2020mthael} \\ \hline
    IoT, IoMT & Google Play & \href{https://play.google.com/}{googleplay.com} & B & \cite{ren2020end}, \cite{amin2022deep} \\ \hline
    
    IoT & Bot-IoT (network data) & \href{https://research.unsw.edu.au/projects/bot-iot-dataset}{The Bot-IoT Dataset | UNSW Research} & M \& B & \cite{guizani2020network} \\ \hline
    
    IoT & Linux system files & \href{https://www.kernel.org/}{/bin /user/bin /sbin} & B & \cite{shobana2020novel},\cite{chaganti2022deep} \\ \hline
    
    IoT & IoT-23 (network traffic data) & \href{https://www.stratosphereips.org/datasets-iot23}{IoT-23 Dataset: A labeled dataset of Malware and Benign IoT Traffic} & M \& B & \cite{nobakht2023demd}, \cite{jeon2021hybrid} \\ \hline
    
    IoT, IIoT & Derbin & \href{https://www.sec.cs.tu-bs.de/~danarp/drebin/}{sec.cs.tu-bs.de/~danarp/drebin/} & M & \cite{yuan2021iot},\cite{karbab2018maldozer}, \cite{yumlembam2022iot}, \cite{aamir2024amddlmodel}, 
    \cite{khowaja2021q}\\ \hline
        
    IoT, IoMT & IoT malware & \href{https://www.kaggle.com/datasets/anaselmasry/iot-malware}{kaggle.com/dgawlik/3s19-iot-malware} & M \& B & \cite{asam2022iot}, \cite{yuan2021iot}, \cite{khan2021hybrid} \\ \hline
    
    IoT, IoMT & Big 2015  & \href{https://www.kaggle.com/c/malware-classification}{kaggle.com/c/malware-classification} & M & \cite{akhtar2022detection}, \cite{sharma2023windows}, \cite{yuan2021iot}, \cite{ravi2022attention} \\ \hline

    IIoT & Big 2015  & \href{https://github.com/mohammadreza-babaeimosleh/}{github.com/mohammadreza-babaeimosleh/} & M & \cite{mosleh2024efficient} \\ \hline
    
    IoT & Malimg  & \href{https://www.kaggle.com/code/gvyshnya/malimg-coatnet-model}{kaggle.com/datasets/satya11/malimg-dataset} & M & \cite{sharma2023windows} \\ \hline

    IIoT & Malimg  & \href{https://vision.ece.ucsb.edu/research/signal-processing-malware-analysis}{vision.ece.ucsb.edu/research/signal-processing-malware-analysis} & M & \cite{naeem2020malware}, \cite{ahmed2022multilayer},\cite{kim2022iiot}, \cite{smmarwar2023ai}, \cite{kumar2024image}\\ \hline
    
    IoT & theZoo & \href{https://github.com/ytisf/theZoo}{github.com/ytisf/theZoo} & M \& B & \cite{sharma2023windows} \\ \hline
    
    IoT & VXHeaven & \href{https://vx-underground.org/archive/VxHeaven/index.html}{vx-underground.org/archive/VxHeaven/index.html} & M \& B & \cite{sharma2023windows} \\ \hline
    
    IoT & Malgenome & \href{http://www.malgenomeproject.org/}{malgenomeproject.org} & M & \cite{karbab2018maldozer} \\ \hline
    
    IoT & CICAndMal2017 & \href{https://www.unb.ca/cic/datasets/android-adware.html}{unb.ca/cic/datasets/android-adware.html} & M \& B & \cite{waqar2022malware} \\ \hline
    
    IoT, IoMT & CICMalDroid2020 & \href{https://www.unb.ca/cic/datasets/maldroid-2020.html}{unb.ca/cic/datasets/maldroid-2020.html} & M \& B & \cite{yumlembam2022iot}, \cite{nawshin2024ai}\\ \hline
    
    IoT & CICInvesAndMal2019  & \href{https://www.unb.ca/cic/datasets/invesandmal2019.html}{unb.ca/cic/datasets/invesigating-android-malware-2019.html} & M \& B & \cite{ksibi2024efficient} \\ \hline
    
    IoT & CyberIOCs & \href{https://www.cyberiocs.com/}{cyberiocs.com} & M & \cite{alasmary2019analyzing} \\ \hline

    IIoT & Collected by Azmoodeh et al & \href{https://github.com/CyberScienceLab/Our-Papers}{github.com/CyberScienceLab/Our-Papers} & M \& B & \cite{esmaeili2022iiot} \\ \hline
        
    IIoT & Playdrone & \href{https://archive.org/details/playdrone-apks}{archive.org/details/playdrone-apks} & B & \cite{khowaja2021q} \\ \hline
    
    IoBT & Collected from VirusTotal and official IoT app stores & \href{https://github.com/azmoodeh/IoTMalwareDetection}{github.com/azmoodeh/IoTMalwareDetection} & M \& B & \cite{azmoodeh2018robust},\cite{shah2023heucrip} \\ \hline
    
    IoMT & NSL-KDD & \href{https://kdd.ics.uci.edu/databases/kddcup99/kddcup99.html}{kdd.ics.uci.edu/databases/kddcup99/kddcup99.html} & M \& B & \cite{saheed2021efficient} \\ \hline
    
    IoMT & CCCS-CIC-AndMal-2020  & \href{https://www.unb.ca/cic/datasets/andmal2020.html}{unb.ca/cic/datasets/cccs-cic-andmal-2020.html} & M \& B & \cite{nawshin2024ai} \\ \hline

    IoMT & CDMC-2020-IoMT-Malware  & \href{https://www.csmining.org/CyberAICup2024/home.html}{CDMC-2020-IoMT-Malware} & M \& B & \cite{ravi2022attention} \\ \hline

    IoV & Combination of CIC DoS 2016, CICIDS 2017, CSE-CIC-IDS 2018 (DDoS and benign traffic) & \href{https://www.kaggle.com/datasets/devendra416/ddos-datasets}{kaggle.com/c/dataset-ddos2018} & M \& B & \cite{ullah2022hdl} \\ \hline
    
    IoV & Car-Hacking (network traffic) & \href{https://ocslab.hksecurity.net/Datasets/CAN-intrusion-dataset}{hksecurity.net/car-hacking-dataset/} & M \& B & \cite{ullah2022hdl}, \cite{ashraf2020novel}, \cite{song2020vehicle} \\ \hline
    
    IoV & UNSW-NB15 (network traffic) & \href{https://research.unsw.edu.au/projects/unsw-nb15-dataset}{research.unsw.edu.au/projects/unsw-nb15-dataset} & M \& B & \cite{ashraf2020novel}\\ \hline
    \end{tabular}
    \vspace{1mm}
    \label{tab:dataset}
\end{table*}

\begin{table*}[!ht]
    \centering
    \def\arraystretch{1.5}
    \caption{Summary of DL-based IoT malware analysis. Approach S, D, and H represent static, dynamic and hybrid, respectively. Model performance denotes A as Accuracy, P as Precision, R as Recall and F as F1-Score. Note that N/A indicates not addressed.}
    \rowcolors{2}{gray!25}{white}
    \begin{tabular}{|p{1.1cm}|p{6.1cm}|p{1.3cm}|p{1.7cm}|p{0.9cm}
    |p{2.4cm}|}
    \rowcolor{gray!25}
    \hline
    \textbf{Reference} & \textbf{Main Contribution} & \textbf{OS \& Arch.} & \textbf{Approach \& Features} & \textbf{Model} 
    & \textbf{Model Performance} \\
    \hline
    Haddad Pajouh et al (2018). \cite{haddadpajouh2018deep} & 
    Using LSTM-based RNN for detecting IoT malware through Opcode sequence analysis & 
    OS: Linux &
    S (OpCode) &
    LSTM & 
    \begin{minipage}[t]{\linewidth}
        \begin{itemize}[leftmargin=*]
        \item A: 98.18
        \end{itemize}
        \vspace{1mm}
    \end{minipage}
    \\
    \hline
    Vasan et al (2020). \cite{vasan2020mthael} & 
      Detecting obfuscated IoT malware, including metamorphic and polymorphic by using RNN-CNN with heterogeneous features & 
      OS: N/A, Arch: ARM, Intel & 
      S (OpCode) & 
      RNN-CNN & 

      \begin{minipage}[t]{\linewidth}
      For ARM:
     \begin{itemize}[leftmargin=*]     
         \item A: 99.98, P: 99.96, R: 99.97, F: 99.94
     \end{itemize}
     \vspace{1mm}
     \end{minipage}
     \\
     \hline
      Guizani et al (2020). \cite{guizani2020network} & 
      Developing NFV system integrated with an RNN-LSTM model for detecting malware in large-scale IoT networks & 
      N/A & 
      D (network traffic) & 
       RNN-LSTM & 
       \begin{minipage}[t]{\linewidth}
       \begin{itemize}[leftmargin=*]     
          \item Avg. A: 85  
     \end{itemize}
     \vspace{1mm}
     \end{minipage}
     \\
     \hline

     Shobana et al (2020). \cite{shobana2020novel} & 
    Retrieving system calls information by applying N-grams and vectorization techniques to develop RNN-based IoT malware detection model  & 
    OS: Ubuntu &
    Hybrid: System calls  & 
    RNN & 
     \begin{minipage}[t]{\linewidth}
        \begin{itemize}[leftmargin=*]
        \item A: 98.712
        \end{itemize} 
     \vspace{1mm}
     \end{minipage}
    \\
     \hline

Jeon et al (2020). \cite{jeon2020dynamic} & 
Detecting both known and new IoT malware by dynamically analyzing malware behavior in a cloud-based environment and using CNN  & 
OS: Embedded Linux & 
D (network traffic, sys calls, vfs, and processes data)  & 
CNN & 
     \begin{minipage}[t]{\linewidth}
        \begin{itemize}[leftmargin=*]
           \item A: 99.28
        \end{itemize}
     \vspace{1mm}
     \end{minipage}
           \\
     \hline
Dib et al (2021).\cite{dib2021multi} & 
Developing a multi-level approach for static IoT malware detection using CNN and LSTM  & 
OS: Linux & 
S (ELF binaries)  & 
CNN, LSTM & 

     \begin{minipage}[t]{\linewidth}
        \begin{itemize}[leftmargin=*]
        \item A: 99.78, F: 99.57
        \end{itemize} 
     \vspace{1mm}
     \end{minipage} \\
     \hline
    Nobakht et al (2023).\cite{nobakht2023demd} & 
    Proposing DEMD-IoT for IoT malware detection using CNNs and Random Forest meta-learner  & 
    N/A & 
    D (network traffic)  & 
    1D-CNNs, RF & 
    \begin{minipage}[t]{\linewidth}
        \begin{itemize}[leftmargin=*]
           \item A: 99.9, P: 99.83, R: 99.97, F: 99.9
        \end{itemize}
    \vspace{1mm}
    \end{minipage}
    \\
    \hline

    \end{tabular}
    \vspace{1mm}
    \label{tab:iot}
\end{table*}

\begin{table*}[!t]
    \centering
    \def\arraystretch{1.5}
    \ContinuedFloat
    \caption{Summary of DL-based IoT malware analysis (continued).}
    \rowcolors{2}{gray!25}{white}
    \begin{tabular}{|p{1.1cm}|p{4.8cm}|p{1.3cm}|p{1.7cm}|p{0.9cm}
    |p{3.7cm}|}
    \rowcolor{gray!25}
    \hline
    \textbf{Reference} & \textbf{Main Contribution} & \textbf{OS \& Arch.} & \textbf{Approach \& Features} & \textbf{Model} 
    & \textbf{Model Performance} \\
    \hline
  Yuan et al (2021). \cite{yuan2021iot} & 
    Proposing an approach considering the computational resource limitations of IoT devices  & 
    OS: Linux, Android, and Windows& 
    S (Linux, Android, and Windows binaries) & 
    LCNN & 


    \begin{minipage}[t]{\linewidth}
    Linux (ARM) A: 97.162, Android (APP) A: 97.413, Windows A: 99.356
     \vspace{1mm}
     \end{minipage}

    \\
    \hline
    
    Asam et al (2022).\cite{asam2022iot} & 
    Proposing a CNN-based architecture for IoT malware detection, featuring mechanisms for channel boosting and squeezing& 
    OS: Linux & 
    S (ELF binaries) & 
    Channel boosted and squeezed CNN & 
     \begin{minipage}[t]{\linewidth}
     \begin{itemize}[leftmargin=*]     
        \item A: 97.93, P: 98.64, R: 88.73, F: 93.94
     \end{itemize}
     \vspace{1mm}
     \end{minipage}
    \\
    \hline
    Jeon et al (2021). \cite{jeon2021hybrid} & 
    Proposing the HyMalD scheme, which utilizes hybrid analysis combining Bi-LSTM and SPP-Net models to detect a wide variety of IoT malware &
    OS: Windows& 
    H (OpCode, network traffic, processes data, registry, and API calls) &
    Bi-LSTM, SPP-net &

     \begin{minipage}[t]{\linewidth}
     \begin{itemize}[leftmargin=*]     
        \item A: 92.5
     \end{itemize}
     \vspace{1mm}
     \end{minipage} 
    \\
    \hline

    Akhtar et al (2022). \cite{akhtar2022detection} & 
    Developing an application combining the CNN-LSTM models for real-time malware detection & 
    N/A & 
    S (n-gram API) & 
    CNN-LSTM & 
     \begin{minipage}[t]{\linewidth}
     \begin{itemize}[leftmargin=*]     
        \item A: 99, P: 99, R: 99, F: 100
     \end{itemize}
     \vspace{1mm}
     \end{minipage}
     \\
     \hline
    Chaganti et al (2022). \cite{chaganti2022deep} & 
    Developing Cross-architecture IoT malware detection and classification by using the Bi-GRU-CNN deep learning approach & 
    Arch: MIPS, ARM, X68, SuperH4, PPC &
    S (ELF binary) & 
    Bi-GRU-CNN & 

     \begin{minipage}[t]{\linewidth}
     \begin{itemize}[leftmargin=*]     
        \item Detection A: 100
        \item Classification A: 98
     \end{itemize}
     \vspace{1mm}
     \end{minipage}
     \\
     \hline
    Devi et al (2023). \cite{devi2023enhancement} & 
    Developing a deep LSTM-based IoT malware detection approach with enhanced data transmission security using improved Elliptic Curve Cryptography &
    N/A & 
    O (Network traffic) & 
    Deep LSTM & 
    \begin{minipage}[t]{\linewidth}
     \begin{itemize}[leftmargin=*]     
        \item A: 95, P: 92, R: 90
     \end{itemize}
     \vspace{1mm}
     \end{minipage}
    \\
    \hline
   
    Sharma et al (2023). \cite{sharma2023windows}& 
    Developing a malware recognition system that integrates visualization, automatic feature extraction, and classification using traditional and transfer learning &
    OS: Windows &
    S (executable binaries) &
    Custom CNN and Xception &
    \begin{minipage}[t]{\linewidth}
    \textbf{Win - Custom CNN:} A: 98.91, F: 97.54 \\
    \textbf{Win - Xception CNN:} A: 99.20, F: 98.86 \\
    \textbf{IoT - Custom CNN:} A: 99.05, F: 99.20 \\
    \textbf{IoT - Xception CNN:} A: 99.18, F: 98.91

     \vspace{1mm}
     \end{minipage}
     \\
     \hline
    
    \end{tabular}
    \vspace{1mm}
    \label{tab:iot}
\end{table*}

\begin{table*}[!t]
    \centering
    \def\arraystretch{1.5}
    \ContinuedFloat
    \caption{Summary of DL-based IoT malware analysis (continued).}
    \rowcolors{2}{gray!25}{white}
    \begin{tabular}{|p{1.1cm}|p{5.1cm}|p{1.3cm}|p{1.7cm}|p{0.9cm}
    |p{3.4cm}|}
    \rowcolor{gray!25}
    \hline
    \textbf{Reference} & \textbf{Main Contribution} & \textbf{OS \& Arch.} & \textbf{Approach \& Features} & \textbf{Model} 
    & \textbf{Model Performance} \\
    \hline
    Ding et al (2020). \cite{ding2020deeppower} & 
    Developing effective data preprocessing methods to minimize noise in power signals, combined with an attention-based Seq2Seq model for fine-grained analysis of power side-channel signals &
    OS: Linux & 
    D (power signals) & 
    Attention-based Seq2Seq model with LSTM &
    \begin{minipage}[t]{\linewidth}
     \begin{itemize}[leftmargin=*]     
        \item Avg A:  90.4 
     \end{itemize}
     \vspace{1mm}
     \end{minipage}
    \\
    \hline
    
    Karbab et al (2018). \cite{karbab2018maldozer}  & 
    Introducing the MalDozer framework, which utilizes deep learning for Android malware detection and family attribution, along with an automatic feature extraction technique based on sequences of API method calls
    & 
    OS: Android & 
    S (API calls) & 
    CNN & 
  
   
     \begin{minipage}[t]{\linewidth}
\textbf{Malgenome (10-Fold):} P: 99.85, R: 99.85, F: 99.84 \\
    \textbf{Drebin (10-Fold):} P: 99.21, R: 99.21, F: 99.21 \\
    \textbf{MalDozer (10-Fold):} P: 98.18, R: 98.18, F: 98.18

     \vspace{1mm}
     \end{minipage}
    \\
    \hline
    Waqar et al (2022). \cite{waqar2022malware} & 
    Introducing a hybrid deep learning model that combines DNN, BLSTM, and GRU for real-time multiclass malware detection in Android IoT devices &
    OS: Android  & 
    D (API calls and network traffic) &
    DNN-BiLSTM-GRU &

     \begin{minipage}[t]{\linewidth}
     \begin{itemize}[leftmargin=*]  
        \item A: 99.87, P: 99.94, R: 99.9, F: 99.9
     
     \end{itemize}
     \vspace{1mm}
     \end{minipage} 
    \\
    \hline

    Ren et al (2020). \cite{ren2020end} & 
    Developing DexCNN and DexCRNN, two deep learning models for end-to-end Android malware detection, with DexCNN using CNNs and DexCRNN combining CNNs with LSTM and GRU &
    OS: Android & 
    S (bytecode) & 
    DexCNN and DexCRNN & 
     \begin{minipage}[t]{\linewidth}
    \textbf{DexCNN:}
    \begin{itemize}[leftmargin=*] 
        \item A: 93.4, P: 89.7, R: 98.1, F: 93.7

    \end{itemize}
    \textbf{DexCRNN:}
    \begin{itemize}[leftmargin=*]
        \item A: 95.8, P: 95.4, R: 96.2, F: 95.8

    \end{itemize}
     \vspace{1mm}
     \end{minipage}
     \\
     \hline
    Yumlembam et al (2022). \cite{yumlembam2022iot} & 
     Introducing a GNN-based classifier for Android malware detection, utilizing API graph embeddings and Permission, and intent features 
     & 
    OS: Android &
    S (API calls, Permissions, and Intents) & 
    CNN &

    \begin{minipage}[t]{\linewidth}
    \textbf{On CICMaldroid:} A: 98.33, P: 99.18, R: 98.6, F: 98.89 \\
    \textbf{On Drebin:} A: 98.68, P: 95.27, R: 91.08, F: 93.13
     \vspace{1mm}
     \end{minipage}
     \\
     \hline
    Ksibi et al (2024). \cite{ksibi2024efficient} & 
    Introducing a novel CNN-based malware detection approach using pre-trained models like DenseNet169, Xception, InceptionV3, ResNet50, and VGG16 to extract features from Android APK files converted to binary codes and RGB images & 
    OS: Android & 
    S (APK files) & 
Custom CNN models

      
       &
    \begin{minipage}[t]{\linewidth}
    \textbf{ DenseNet169:}
    \begin{itemize}[leftmargin=*] 
        \item A: 95.24
    \end{itemize}
    \textbf{ InceptionV3:}
    \begin{itemize}[leftmargin=*]
         \item A: 95.24  
    \end{itemize}
    \textbf{VGG16:}
    \begin{itemize}[leftmargin=*]
        \item A: 95.83       
    \end{itemize}
     \vspace{1mm}
     \end{minipage}
    \\
    \hline
    
    Aamir et al (2024). \cite{aamir2024amddlmodel}  & 
    Introducing the AMDDL model, a CNN-based deep learning technique for Android malware detection, evaluated with different parameters, filter sizes, epochs, learning rates, and layers &
    OS: Android & 
    S (OpCode, API calls, and Permissions) & 
    CNN & 
  
     \begin{minipage}[t]{\linewidth}
    \begin{itemize}[leftmargin=*] 
        \item A: 99.92, P: 98.61, R: 99.16, F: 98.88

    \end{itemize}
     \vspace{1mm}
     \end{minipage}
    \\
    \hline
    Alasmary et al (2019). \cite{alasmary2019analyzing} & 
    Analyzing the CFG of IoT and Android malware to extract graph-theoretic features. These features are then used to build a deep learning-based detection model & 
    OS: Linux  & 
    S (ELF executable) &
    CNN &


    

     \begin{minipage}[t]{\linewidth}
     \textbf{Malware Detection:} A: 99.66 \\
    \textbf{Malware Classification:} A: 99.32

     \vspace{1mm}
     \end{minipage} 
    \\
    \hline

    \end{tabular}      
    \vspace{1mm}
    \label{tab:iot}
\end{table*}
    

\textbf{LSTM-based approaches for IoT malware detection}. Several studies focus on using LSTM models for detecting IoT malware. HaddadPajouh et al. \cite{haddadpajouh2018deep} applied LSTM to analyze OpCode sequences in ARM-based IoT applications. They collected a dataset of 281 malware and 270 benign samples, extracting and analyzing OpCode sequences to distinguish between malware and benign software. Information Gain (IG) was employed for feature selection to identify the most significant OpCodes. Various configurations of the LSTM model were explored, with the most effective configuration featuring two LSTM layers. The model was assessed using 100 IoT malware samples not seen during training, and the second LSTM configuration demonstrated the highest accuracy at 98.18\%. Similarly, Ding et al. \cite{ding2020deeppower} introduced the DeepPower model, employing an attention-based Seq2Seq architecture with LSTM networks for non-intrusive malware detection via power side-channel analysis. This model processed power signals from various IoT devices infected with real-world malware such as Mirai, Lizkebab, BASHLITE, Tsunami, and BASH Script. DeepPower integrated mel-scaled spectrograms as feature representations, ensuring high-quality input after preprocessing power signals with wavelet denoising and simple moving average filtering. The model achieved an average detection accuracy of 90.4\% across three IoT devices, significantly outperforming existing methods like WattsUpDoc, with a true positive rate of 92.7\% and a false positive rate of 2.9\% for detecting Mirai malware.

\textbf{Combining static and dynamic analysis techniques}. Research by Jeon et al. \cite{jeon2021hybrid} and Devi et al. \cite{devi2023enhancement} demonstrates the integration of static and dynamic analysis for more robust IoT malware detection. Jeon et al. developed the HyMalD framework, combining Bi-LSTM for static feature analysis with SPP-Net for dynamic API call sequence analysis in a virtual environment. Bi-LSTM was utilized to analyze static features extracted from the opcode sequences of IoT device applications, benefiting from its ability to comprehend the context in both directions of the input sequence. For dynamic analysis, SPP-Net was used to examine behavior related to API call sequences dynamically extracted during the execution of potentially malicious files. These API call sequences were transformed into colored images, enabling efficient malware detection through deep learning. HyMalD's effectiveness was evaluated on a dataset comprising both malware and benign files, achieving an accuracy of 92.5\% on 38,166 static files and 14,743 dynamic files, including obfuscated and unobfuscated files. Devi et al. focused on distinguishing attack nodes from normal nodes using trust values derived from contextual features. An Improved Elliptic Curve Cryptography (IECC) algorithm was proposed to prevent detected malware from compromising data transmission among IoT devices. Following the identification of attack nodes, preprocessing and feature extraction strategies, including Linear Discriminant Analysis (LDA), were employed to minimize dimensionality and focus on relevant features. A specially designed Deep LSTM classifier was proposed, achieving 95\% accuracy, 92\% precision, and 90\% recall in detecting malware.

\textbf{CNN-based models for IoT malware detection}. Multiple works have explored CNN-based approaches for detecting IoT malware. Jeon et al. \cite{jeon2020dynamic} used CNNs, specifically ZFNet, to analyze behavioral features extracted from IoT malware during dynamic analysis. This methodology involved executing IoT malware within a controlled, cloud-based virtual environment and extracting and preprocessing features like memory usage, network activities, system calls, and processes into colored images. These images were then used to train the CNN model to differentiate between malware and benign software. The study used 1,000 IoT malware samples and 401 benign files, achieving an accuracy of 99.29\%. Similarly, Asam et al. \cite{asam2022iot} developed the iMDA model, which leverages CNNs for detecting malware in ELF files in Linux-based IoT systems. Their architecture incorporated advanced feature learning techniques such as edge exploration, smoothing, multi-path dilated convolutional operations, and channel squeezing and boosting within the CNN. The iMDA model was evaluated on a Kaggle benchmark IoT dataset containing 14,733 grayscale images of malware and 2,486 grayscale images of benign software, achieving an accuracy of 97.93\%, precision of 98.64\%, recall of 88.73\%, and an F1-score of 93.94\%.

\textbf{Ensemble learning and hybrid models}. The integration of deep learning with ensemble learning techniques has been explored in works like Nobakht et al. \cite{nobakht2023demd} and Vasan et al. \cite{vasan2020mthael}. Nobakht et al. proposed the DEMD-IoT model, combining three distinct one-dimensional CNNs to analyze and learn from IoT network traffic data, enabling the capture of a wide range of patterns indicative of malware. The outputs of these CNNs were integrated using a meta-learner with the Random Forest algorithm to improve predictions. The model was evaluated using the IoT-23 dataset, which includes both benign and malicious network traffic from IoT devices, achieving 99.9\% accuracy, 99.83\% precision, 99.97\% recall, and a 99.9\% F1-score. Vasan et al. introduced the MTHAEL model, which integrates RNN and CNN to detect obfuscated metamorphic and polymorphic IoT malware. The model employs heterogeneous feature selection techniques, such as Information Gain (IG) and OpCode Dictionary, to extract OpCode sequences from IoT applications. The effectiveness of MTHAEL was evaluated on different hardware architectures, including ARM-based CPU, Core-i3, Core-i5, and Dual-core, achieving high effectiveness in malware detection with 99.98\% accuracy, 99.94\% F1-score, 99.96\% precision, and 99.97\% recall for ARM-based architecture.

\textbf{Lightweight and resource-efficient CNN models}. For IoT devices with constrained computational resources, lightweight CNN models have been proposed. Yuan et al. \cite{yuan2021iot} developed an LCNN model designed to be small and efficient while maintaining high accuracy in malware detection across various IoT platforms. This approach involves transforming malware binaries into multidimensional Markov images, which are subsequently classified by an LCNN that incorporates depthwise convolution and channel shuffle operations to significantly reduce the model's size and complexity. The LCNN model was tested on datasets including Linux-based architectures like ARM, MIPS, PPC, Android-based architecture like APP, and Windows-based. The datasets included 25,004 Linux malware samples, 4,020 Android malware samples, and 10,868 Windows malware samples. The accuracy achieved for Linux malware was 97.162\% for ARM, 95.845\% for MIPS, 96.807\% for PPC, 97.413\% for Android APP, and 99.356\% for Windows malware.

\textbf{Deep learning for Android malware detection in IoT environments}. Several studies have focused on using CNNs and RNNs for Android malware detection in IoT environments. Karbab et al. \cite{karbab2018maldozer} developed MalDozer, a CNN-based model that leverages sequences of API method calls extracted from the DEX files of Android applications. The model is trained on different datasets, including Malgenome, Drebin, and a custom MalDozer dataset, comprising over 33,000 malware samples and 37,000 benign apps from sources like Google Play. The feature representation involves converting API method sequences into fixed-length high-dimensional vectors using word embedding techniques such as word2vec and GloVe. MalDozer achieves detection performance with F1-Scores ranging from 96\% to 99\% and false positive rates between 0.06\% and 2\% across different datasets with ranges between 2-Fold and 10-Fold. Ren et al. \cite{ren2020end} introduced the DexCRNN framework, combining CNN with RNN (LSTM and GRU) layers to process raw bytecodes from the classes.dex files of Android applications. The model is trained on a dataset containing 8,000 benign and 8,000 malicious APKs from VirusShare and Google Play Store. The feature representation involves resampling the bytecode sequences into consistent lengths, ensuring efficient processing by the deep learning model. DexCRNN achieves a detection accuracy of 95.8\%, precision of 95.4\%, recall of 96.2\%, and an F1-score of 95.8\%.

\subsection{\textbf{Industrial IoT (IIoT)}}

This subsection focuses on key studies that employ deep learning techniques in the IIoT domain for malware analysis. Table~\ref{tab:iiot} summarizes the research done on DL-based IIoT malware analysis.

\begin{table*}[!t]
    \centering
    \def\arraystretch{1.5}
    \caption{Summary of DL-based IIoT malware analysis. Approach S, D, and H represent static, dynamic and hybrid, respectively. Model performance denotes A as Accuracy, P as Precision, R as Recall and F as F1-Score. Note that N/A indicates not addressed.}
    \rowcolors{2}{gray!25}{white}
    \begin{tabular}{|p{1.1cm}|p{5.1cm}|p{1.3cm}|p{1.7cm}|p{0.9cm}
    |p{3.4cm}|}
    \rowcolor{gray!25}
    \hline
    \textbf{Reference} & \textbf{Main Contribution} & \textbf{OS \& Arch.} & \textbf{Approach \& Features} & \textbf{Model} 
    & \textbf{Model Performance} \\
    \hline
    Naeem et al (2020). \cite{naeem2020malware} & 
    Proposing an architecture for detecting malware attacks in the IIoT environment, utilizing a methodology that integrates malware visualization with a deep CNN model  & 
    OS: Android and Windows  & 
    S (Android APK and Windows executables) & 
     Deep CNN & 
     
     \begin{minipage}[t]{\linewidth}
     \textbf{Leopard Mobile dataset:} A: 97.81, P: 95.16, R: 95.1, F: 95.13 \\
    \textbf{Malimg dataset:} A: 98.74, P: 98.47, R: 98.47, F: 98.46

    \vspace{1mm}
    \end{minipage}
    \\
    \hline
    Ahmed et al (2022). \cite{ahmed2022multilayer} & 
    Developing a 5G-enabled CNN-based system for efficient IIoT malware classification using image representation &
    N/A & 
    S (malware binaries) & 
    CNN & 
      \begin{minipage}[t]{\linewidth}
     \begin{itemize}[leftmargin=*]     
         \item A: 97   
     \end{itemize}
     \vspace{1mm}
     \end{minipage}
    \\
    \hline
    Kim et al (2022). \cite{kim2022iiot}& 
    Developing an edge computing-based malware detection system for smart factories using a CNN model and image visualization technology & 
    N/A & 
    S (malware binaries) & 
    CNN & 
       \begin{minipage}[t]{\linewidth}
       \begin{itemize}[leftmargin=*]     
          \item A: 98.93, P: 98.93, R: 98.93, F: 98.92
    \end{itemize}
    \vspace{1mm}
    \end{minipage}
    \\
    \hline
    Esmaeili et al (2022). \cite{esmaeili2022iiot}& 
    Developing a stateful query analysis (SQA) approach to detect and prevent query-based black-box adversarial attacks on IIoT systems by analyzing query sequences &
    Arch: ARM & 
    S (byte code) & 
    CNN & 
    \begin{minipage}[t]{\linewidth}
        \begin{itemize}[leftmargin=*]
        \item Detection rate: 93.1
        \end{itemize} 
     \vspace{1mm}
     \end{minipage}
    \\
    \hline  

    Mosleh et al (2024). \cite{mosleh2024efficient} & 
    Introducing HierarchicalCloudDNN, a distributed framework using deep learning to scale malware detection from IIoT devices to the edge and cloud, enhancing speed and reducing resource use &
    N/A & 
    S (examines the firmware of devices) & 
Compact CNN models &
    \begin{minipage}[t]{\linewidth}
     \begin{itemize}[leftmargin=*]     
        \item A: 98.90, P: 98.76, R: 96.88, F: 97.81
     \end{itemize}
     \vspace{1mm}
     \end{minipage}
    \\
    \hline

    Smmarwar et al (2023). \cite{smmarwar2023ai} & 
    Introducing the D3WT-CNN-LSTM model, combining Double-Density DWT for feature extraction with a hybrid CNN-LSTM for malware detection and classification &
    N/A & 
    S (OpCodes, strings, and byte codes) & 
    CNN-LSTM &
   


       \begin{minipage}[t]{\linewidth}
       \textbf{IoT malware:} A: 99.98, P: 99.67, R: 99.68, F: 99.69 \\
   \textbf{MMB-15:} A: 96.97, P: 97.62, R: 97.92, F: 97.93 \\
   \textbf{Malimg:} A: 99.96, P: 99.81, R: 99.96, F: 99.97


     \vspace{1mm}
     \end{minipage}
    \\
    \hline
Kumar et al (2024). \cite{kumar2024image} & 
     Introducing CNN-AE, a model that combines CNN with a two-level Autoencoder for malware detection by transforming binary programs into gray-scale images, extracting and reducing features, and classifying with machine learning and deep learning &
    N/A & 
    H (binary executable and network traffic)  & 
    \begin{minipage}[t]{\linewidth}
    \textbf{For feature extraction:} Compact CNN models \\
    \textbf{For classification:} MLP

       \vspace{1mm}
       \end{minipage} &

 \begin{minipage}[t]{\linewidth}
     \textbf{Best MLP performance achieved by using VGG16 model:}
      \begin{itemize}[leftmargin=*]
         \item  A: 98.55
         \item Weighted P: 98.60
         \item Weighted R: 98.60
         \item Weighted F: 98.60
         \end{itemize}
       \vspace{1mm}
       \end{minipage}

         \\
    \hline

    Khowaja et al (2021).~\cite{khowaja2021q} & 
     Introducing a model that combines Q-learning with active learning, PSE, SAE, and LSTM, enabling efficient malware classification with fewer labeled examples &
    OS: Android, Windows, and IOS & 
    H (Permissions, ICC, API calls, dynamic mobile security features) &
    LSTM with Q-learning &



 \begin{minipage}[t]{\linewidth}
  \textbf{Using 50\% training data:} \\
    Classification A: 95.1, Adversarial: 86.9
       \vspace{1mm}
       \end{minipage}

    \\
    \hline

    \end{tabular}
    \vspace{1mm}
    \label{tab:iiot}
\end{table*}

\textbf{DCNN-based approaches for IIoT malware detection}. Deep Convolutional Neural Network (DCNN) techniques are leveraged across several reviewed works in the IIoT domain to analyze malware. Naeem et al. \cite{naeem2020malware} proposed a method that integrates malware visualization with DCNN for more comprehensive malware analysis. This involves transforming binary files from IIoT applications into color images, which are then used as input for the DCNN model. The architecture is tailored to process these visualized binary files, extracting important features for malware detection. The approach was tested using two datasets: the Leopard Mobile dataset, comprising 14,733 malware and 2,486 benign samples, and the Malimg dataset, containing 9,339 samples from 25 malware families. For the Leopard Mobile dataset, optimal results were obtained with an image ratio of 224x224, showing an accuracy of 97.81\%, precision of 95.16\%, recall of 95.10\%, and an F1-Score of 95.13\%. On the other hand, the Malimg dataset displayed higher metrics, with an accuracy of 98.74\%, precision of 98.47\%, recall of 98.47\%, and an F1-Score of 98.46\%.

Ahmed et al. \cite{ahmed2022multilayer} introduced a deep learning framework to enhance malware detection within the 5G-enabled IIoT domain. This framework involves converting malware binaries into 2-D grayscale images of 64x64 pixels and applying a multilayer CNN model for classification. The model includes several convolutional layers with ReLU activation, max-pooling layers for dimensionality reduction, and fully connected layers for classification, utilizing a softmax function in the final layer. Tested with the Malimg dataset containing 9,339 malware samples, this method achieved an accuracy of 97\%.

\textbf{CNN-based models for IIoT malware detection}. CNN architectures have been designed for IIoT environments such as smart factories. Kim et al. \cite{kim2022iiot} present a CNN architecture specifically designed for malware detection within smart factories, processing malware binaries into grayscale images of 112x112 pixels. The model is trained over 40 epochs, incorporating a dropout rate of 0.45 to prevent overfitting, and using a batch size of 32. Utilizing the Malimg dataset, this approach achieves an accuracy of 98.93\%, precision of 98.93\%, recall of 98.93\%, and an F1-Score of 98.92\%.

Kumar et al. \cite{kumar2024image} present a hybrid deep learning model named CNN-AE, which utilizes CNN and a two-level Autoencoder for malware detection in IIoT. Combining CNN models with the autoencoder layers provides a fast response and improved accuracy, leveraging deep layers of CNN for better feature extraction. The model is trained on data containing instances of malware from the Malimg dataset, covering various malware families. The feature representation involves transforming binary programs into grayscale images, extracting textural features using different deep CNN architectures, and reducing the dimensions of these features through the two-level autoencoder. The experimental results show that the CNN-AE model (VGG16) achieved a high test accuracy of 98.55\%, with a weighted precision of 98.60\%, recall of 98.60\%, and an F1-Score of 98.60\%.

\textbf{Detecting adversarial attacks in IIoT systems}. Esmaeili et al. \cite{esmaeili2022iiot} focus on detecting query-based black-box adversarial attacks in the IIoT domain. They introduce the Stateful Query Analysis (SQA) approach, utilizing a history-based method to detect potential adversarial attacks through sequences of queries. The SQA methodology comprises a similarity encoder and a classifier, both based on CNNs, with the similarity encoder analyzing sequences of queries for patterns indicative of adversarial intent. This allows for the preemptive identification of adversarial scenarios before successful execution. The data, consisting of bytecode sequences from IIoT applications, is transformed into 2-D grayscale images for CNN analysis, optimizing the model for this task and demonstrating a detection rate of 93.1\% across various adversarial examples. A dataset of ARM-based IIoT malware samples is used in this study.

\textbf{Combining deep learning techniques for IIoT malware detection}. Significant works in the field of combining multiple deep learning-based approaches for IIoT malware analysis are highlighted as follows:

Mosleh et al. \cite{mosleh2024efficient} present a deep learning model named HierarchicalCloudDNN, designed for IIoT malware classification. This model utilizes a combination of modified SqueezeNet, MobileViT, MobileNetV2, and ResNet-18 architectures to achieve high performance across different layers of a hierarchical pipeline. The model is trained on data containing instances of malware from the BIG 2015 dataset, covering nine malware families. The feature representation involves converting firmware (.byte files) into 2D images processed through resizing and normalization steps. The HierarchicalCloudDNN model demonstrated impressive performance, achieving an accuracy of 98.90\%, precision of 98.76\%, recall of 96.88\%, and an F1-Score of 97.81\%.

Smmarwar et al. \cite{smmarwar2023ai} present a hybrid deep learning model named D3WT-CNN-LSTM. This model utilizes Double-Density Discrete Wavelet Transform (D3WT) for feature extraction and combines CNN and LSTM layers for enhanced classification performance. Integrating D3WT with CNN and LSTM layers ensures fast response times and improved detection accuracy by leveraging the strengths of both convolutional and recurrent neural networks. The model is trained on data containing instances of various malware types from three datasets: IoT malware, Microsoft BIG-2015, and Malimg, covering a wide range of threats such as worms, adware, backdoors, Trojans, and obfuscated malware. The feature representation involves converting malware byte files into grayscale images, which are then processed using D3WT to extract approximate and detailed coefficients, and feeding these features into the CNN-LSTM model. For the IoT malware dataset, the D3WT-CNN-LSTM model achieved an accuracy of 99.98\%, while for the MMB-15 dataset, it reached 96.97\%, and for the Malimg dataset, it demonstrated an accuracy of 99.96\%.

Khowaja et al. \cite{khowaja2021q} introduce a deep learning model named Q-LSTM that combines Q-learning and LSTM for malware defense in IIoT. The model integrates phase space embedding (PSE) and sparse autoencoder (SAE) layers for enhanced accuracy and fast response. The model is trained on datasets from Drebin, VirusShare, and Playdrone; it uses static and dynamic features, including ICC features, permissions, and API calls. The results show that even with 25\% training data, Q-LSTM demonstrates better resiliency to adversarial attacks than existing methods, highlighting the effectiveness of the action-value function. Using 50\% training data, the model achieves a classification accuracy of 95.1\% and 86.9\% with adversarial samples.

\subsection{\textbf{Internet of Battlefield Things (IoBT)}}
This subsection explores the research (summarized in Table~\ref{tab:iobt}) on using deep learning techniques in the \begin{table*}[!t]
    \centering
    \def\arraystretch{1.5}
    \caption{Summary of DL-based IoBT malware analysis. Approach S, D, and H represent static, dynamic and hybrid, respectively. Model performance denotes A as Accuracy, P as Precision, R as Recall and F as F1-Score. Note that N/A indicates not addressed.}
    \rowcolors{2}{gray!25}{white}
    \begin{tabular}{|p{1.1cm}|p{6.1cm}|p{1.3cm}|p{1.7cm}|p{0.9cm}
    |p{2.4cm}|}
    \rowcolor{gray!25}
    \hline
    \textbf{Reference} & \textbf{Main Contribution} & \textbf{OS \& Arch.} & \textbf{Approach \& Features} & \textbf{Model} 
    & \textbf{Model Performance} \\
    \hline
    Azmoodeh et al (2018). \cite{azmoodeh2018robust} & 
    Employing a class-wise feature selection to reject less important opcodes to counter junk-code insertion attacks &
    Arch: ARM &
    S (OpCode to OpCode-sequence graph) &
    Deep EigenLearning &
     \begin{minipage}[t]{\linewidth}
     \begin{itemize}[leftmargin=*]
         \item A: 99.68, P: 98.59, R:98.37, F: 98.48
     \end{itemize}
     \vspace{1mm}
     \end{minipage}
     \\
     \hline
    Shah et al (2023). \cite{shah2023heucrip}& 
    Developing HeuCrip, a probability CFG generation technique combining crisp and heuristic methods to enhance malware detection accuracy in IoBT environments &
    Arch: ARM & 
    S (OpCode to HeuCrip graph) &
    CNN & 
     \begin{minipage}[t]{\linewidth}
        \begin{itemize}[leftmargin=*]
            \item A: 99.93, P: 99.62, R: 98.43, F: 98.85 
        \end{itemize}
     \vspace{1mm}
     \end{minipage}
    \\
    \hline    
    
    \end{tabular}
    \vspace{1mm}
    \label{tab:iobt}
\end{table*}
 IoBT field, focusing on their effectiveness in combating malware threats.


Azmoodeh et al. \cite{azmoodeh2018robust} present a deep learning-based method for detecting malware in IoBT devices through the analysis of the device's OpCode sequence. This approach transforms OpCodes into a vector space and employs deep eigenspace learning to distinguish between benign and malicious applications. It demonstrates resilience against junk code insertion attacks, thereby enhancing its robustness in malware detection. The method utilizes Class-Wise Information Gain for feature selection, converting the selected features into a graph representation where nodes represent OpCodes and edges indicate the probability of their occurrences. This graph is then employed in the classification of IoBT malware and benign applications using eigenspace and deep convolutional network techniques, achieving an impressive accuracy rate of 99.68\%, with precision, recall, and F1-Score rates of 98.59\%, 98.37\%, and 98.48\% respectively. The dataset used consists of 1,078 benign and 128 malware samples, all compatible with ARM IoT applications.

Shah et al. \cite{shah2023heucrip} introduce the HeuCrip technique to enhance malware detection, similar to previous work. The HeuCrip method involves disassembling binary files into OpCodes and generating probabilistic control flow graphs (CFGs), where vertices represent OpCodes and edges indicate their occurrence probabilities. HeuCrip transforms these graphs into eigenvectors and eigenvalues, which serve as inputs for a CNN tailored for malware classification. The CNN architecture used in HeuCrip includes an input layer with 166 neurons to handle the feature vectors from the HeuCrip-generated CFGs. It features two hidden layers with 64 and 128 neurons, respectively, both utilizing the ReLU activation function for non-linear transformations, and a sigmoid-activated output layer with 2 neurons for the binary classification task of distinguishing between malware and benign files. The dataset includes 1,089 benign files and 128 malware files, achieving an accuracy of 99.93\%, a precision of 99.62\%, a recall of 98.43\%, and an F1-Score of 98.85\%.

\subsection{\textbf{Internet of Things in Medical Technology (IoMT)}}
This subsection presents an in-depth analysis of how deep learning techniques are being implemented for malware detection in the rapidly evolving IoMT domain, which is summarized in Table~\ref{tab:iomt}.

\begin{table*}[!ht]
    \centering
    \def\arraystretch{1.5}
    \caption{Summary of DL-based IoMT malware analysis. Approach S, D, and H represent static, dynamic and hybrid, respectively. Model performance denotes A as Accuracy, P as Precision, R as Recall and F as F1-Score. Note that N/A indicates not addressed.}
    \rowcolors{2}{gray!25}{white}
    \begin{tabular}{|p{1.1cm}|p{5.1cm}|p{1.3cm}|p{1.7cm}|p{0.9cm}
    |p{3.4cm}|}
    \rowcolor{gray!25}
    \hline
    \textbf{Reference} & \textbf{Main Contribution} & \textbf{OS \& Arch.} & \textbf{Approach \& Features} & \textbf{Model} 
    & \textbf{Model Performance} \\
    \hline
    Saheed et al (2021). \cite{saheed2021efficient} & 
    Employs DRNN and supervised machine learning models utilized to classify and detect cyber threats in the IoMT environment & 
    N/A & 
    H (network traffic) & 
    DRNN & 
     \begin{minipage}[t]{\linewidth}
     \textbf{For DRNN model:}
     \begin{itemize}[leftmargin=*]
         \item A: 96.08, P: 85.63, R: 85.63
     \end{itemize}
     \vspace{1mm}
     \end{minipage}
    \\
    \hline
    Amin et al (2022). \cite{amin2022deep}& 
    Introducing a lightweight, adaptable malware detector for distinguishing between benign and malicious applications in Android smartphones and IoT devices within the IoMT domain &
    OS: Android
     & 
    S (byte code, Prune OpCode, and System call) & 
    FCN and LSTM & 

  
     \begin{minipage}[t]{\linewidth}
      \begin{itemize}[leftmargin=*]
 \item \textbf{\textit{LSTM model:}}
         \item \textbf{Android:} P: 99.5, R: 85.63, F: 99
         \item \textbf{IoT:} P: 99.5, R: 99.5, F: 99
         \item \textbf{\textit{FCN model:}}
         \item \textbf{Android:} P: 98.5, R: 98, F: 97
         \item \textbf{IoT:} P: 99.5, R: 99.5, F: 99
     \end{itemize}
     \vspace{1mm}
     \end{minipage}
     \\
     \hline
    Khan et al (2021). \cite{khan2021hybrid}& 
    Developing an SDN-enabled, hybrid DL framework for detecting sophisticated malware in IoMT ecosystems without burdening devices &
    N/A & 
    S (OpCode) & 
    CNN-LSTM & 
     \begin{minipage}[t]{\linewidth}
      \textbf{On 10-Fold validation:} A: 99.96, P: 99.34, R: 99.11, F: 100

        \vspace{1mm}
     \end{minipage}
     \\
     \hline
    Ravi et al (2022). \cite{ravi2022attention}& 
    Introducing a CNN-Bi-LSTM approach for cross-architecture IoMT malware detection, classification, and CPU architecture identification &
    OS: Linux & 
    S (byte sequence) & 
    CNN-Bi-LSTM & 
     \begin{minipage}[t]{\linewidth}
     \textbf{For IoMT malware detection:} A: 95, P: 96, R: 95, F: 95 \\
    \textbf{For IoMT malware classification:} A: 94, P: 96, R: 87, F: 91
     
     \vspace{1mm}
     \end{minipage}
    \\
    \hline   

    Nawshin et al (2024). \cite{nawshin2024ai}& 
    Introducing DP-RFECV-FNN, a privacy-preserving solution combining Differential Privacy with a Feedforward Neural Network for Android malware detection and classification using static and dynamic features &
    OS: Android  & 
    \begin{minipage}[t]{\linewidth}
     H (activities, intents, services, permissions, metadata, system calls, binder calls, and composite behavior)
       \vspace{1mm}
       \end{minipage} & 
    DP-RFECV-FNN & 
     
     \begin{minipage}[t]{\linewidth}
     \textbf{Static features:} Accuracy: 97.78 to 99.21 \\
    \textbf{Dynamic features:} Accuracy: 93.49 to 94.36
     
        \vspace{1mm}
        \end{minipage}

    \\
    \hline
He et al (2023). \cite{he2023image}& 
    Introducing DRL-HMD, a hybrid framework that converts hardware data into images, uses transfer learning to enhance DNN models, and employs a DRL agent with MLPs to dynamically select the best malware detector & 
    OS: Linux &
    O (HPCs) &
    DNN, DRL & 
    
     \begin{minipage}[t]{\linewidth}
        \begin{itemize}[leftmargin=*]
            \item A: 99, F: 99, AUC: 99
        \end{itemize}
        \vspace{1mm}
         \end{minipage}
    
    \\
    \hline    
    
    \end{tabular}
    \vspace{1mm}
    \label{tab:iomt}
\end{table*}

\begin{table*}[!ht]
    \centering
    \def\arraystretch{1.5}
     \caption{Summary of DL-based IoV malware analysis. Approach S, D, and H represent static, dynamic and hybrid, respectively. Model performance denotes A as Accuracy, P as Precision, R as Recall and F as F1-Score. Note that N/A indicates not addressed.}
    \rowcolors{2}{gray!25}{white}
    \begin{tabular}{|p{1.1cm}|p{5.1cm}|p{1.3cm}|p{1.7cm}|p{0.9cm}
    |p{3.4cm}|}
    \rowcolor{gray!25}
    \hline
    \textbf{Reference} & \textbf{Main Contribution} & \textbf{OS \& Arch.} & \textbf{Approach \& Features} & \textbf{Model} 
    & \textbf{Model Performance} \\
    \hline
    Safi Ullah et al (2022). \cite{ullah2022hdl} & 
     Developing the HDL-IDS model that combines LSTM and GRU to improve the attack detection accuracy and response time for intrusion detection in IoV &
     N/A &
     O (IoV cyber attack data) &
     LSTM, GRU & 

     \begin{minipage}[t]{\linewidth}
      \textbf{DDoS dataset:} A: 99.5, P: 99.5, R: 99.6, F: 99.5 \\
    \textbf{Car-hacking dataset:} A: 99.9, P: 99.9, R: 99.9, F: 99.9

     \vspace{1mm}
     \end{minipage}
     \\
     \hline
    Loukas et al (2017). \cite{loukas2017cloud}& 
    Implementing a deep learning-based intrusion detection system on a robotic vehicle, evaluated against DoS, command injection, and malware attacks, compared with traditional methods, and exploring computational offloading for enhanced detection in resource-limited environments &
    OS: Fedora (Linux distribution) &
    O (network traffic) & 
    MLP and RNN with LSTM & 


     \begin{minipage}[t]{\linewidth}
     \textbf{RNN:}
        \begin{itemize}[leftmargin=*]
            \item Avg. A: 86.9\%
     \end{itemize}
     \vspace{1mm}
     \end{minipage}

     \begin{minipage}[t]{\linewidth}
     \textbf{MLP:}
        \begin{itemize}[leftmargin=*]
            \item Avg. A: 78.3\%
        \end{itemize}
     \vspace{1mm}
     \end{minipage}
    \\
    \hline    

    Nie et al (2020). \cite{nie2020data}& 
    Introducing a CNN-based deep learning architecture to analyze RSU link load behaviors in IoV, detecting intrusions by extracting features from link loads &
    N/A & 
    O (network traffic) & 
    CNN & 
     CNN performs better with 2 timeslots than 7, achieving over 97\% in accuracy, precision, recall, and F1-score for both 12 and 30 OBUs
    \\
    \hline 

    Ashraf et al (2020). \cite{ashraf2020novel}& 
    Presenting a statistical feature extraction technique and an LSTM autoencoder-based IDS for AVs and IoVs, detecting attacks in CAN bus and external networks &
    N/A & 
    O (network traffic) & 
    LSTM AE & 
      \begin{minipage}[t]{\linewidth}
       \textbf{Car hacking dataset:} A: 99, P: 99, R: 100, F: 99 \\
    \textbf{UNSW-NB15 dataset:} A: 96, P: 100, R: 97, F: 98

       \vspace{1mm}
       \end{minipage}
    \\
    \hline 

    Song et al (2020). \cite{song2020vehicle}& 
    Proposing an IDS based on a DCNN, specifically a reduced version of the Inception-ResNet architecture & 
    N/A & 
    O (CAN bus traffic with injected messages) &
    Inception-ResNet DCNN &
   
      \begin{minipage}[t]{\linewidth}
      \textbf{Best performance with RPM spoofing dataset:} \\
    P: 99.99, R: 99.94, F: 99.96, FNR: 7, ER: 4
       \vspace{1mm}
       \end{minipage}

    \\
    \hline 
    
    \end{tabular}
    \vspace{1mm}
    \label{tab:iov}
\end{table*}

\textbf{DRNN and LSTM-based approaches for IoMT malware detection}. Deep RNN (DRNN) and LSTM approaches are employed in significant studies within the IoMT domain to analyze malware. Saheed et al. \cite{saheed2021efficient} introduced a cyber-attack detection framework that utilizes DRNN and Supervised Machine Learning (SML) algorithms, including Random Forest, Decision Tree, KNN, and Ridge Classifier. The process initiates with the preprocessing and normalization of network data, followed by feature optimization using the Particle Swarm Optimization (PSO) algorithm. This optimized dataset is subsequently classified using DRNN and various supervised ML models to identify cyber threats in the IoMT environment. The DRNN technique utilizes the sequential data characteristics, with the model's input derived from network data relevant to cyber attacks such as DoS, brute force, and botnet attacks, utilizing the NSL-KDD dataset for intrusion detection. The DRNN model exhibits an accuracy of 96.08\%, recall of 85.63\%, and precision of 85.63\%. Conversely, the Random Forest model demonstrates an accuracy of 99.76\%, recall of 96.45\%, and precision of 96.75\%.

Amin et al. \cite{amin2022deep} tackle the challenge of protecting Android and IoT devices from malware threats. They present a deep learning-based feature detector for malware detection. The method includes extracting bytecode from Android Package Kits (APKs), pruning opcodes and system calls, and applying one hot encoding to generate a vector representation of OpCode and system call data. A dataset including 11,200 malicious apps from VirusShare and 16,680 benign apps from the Google Play Store and third-party sites like Amazon for the Android dataset is employed. Two distinct architectures are utilized for the classification phase: a fully connected network (FCN) with Softmax activation and an LSTM model. The performance metrics for the Android dataset revealed that the LSTM model achieved precision and recall of 99.5\% and an F1-score of 99\%. Similarly, for the IoT dataset, the LSTM model reported the same high precision and recall of 99.5\% and an F1-score of 99.5\%. Meanwhile, the FCN model displayed a slightly lower performance on the Android dataset with a precision of 98.5\%, recall of 98\%, and F1-score of 97\%.

\textbf{Hybrid deep learning models for IoMT malware detection}. Other studies leverage a combination of deep learning techniques to analyze malware. Khan et al. \cite{khan2021hybrid} present a hybrid DL and Software Defined Networking (SDN)-enabled framework for detecting sophisticated IoMT malware. This approach employs a Hybrid CNN-LSTM architecture designed to function without imposing additional burdens on the resource-constrained IoMT devices. The methodology includes extracting benign features from ELF files and malware from OpCode sequences, using Radare2 for disassembly, and a Bag of Words technique for feature vector creation. The model combines CNN for significant feature extraction and LSTM to learn feature interdependence, addressing long-term dependency issues. Utilizing a dataset of 128 malware samples and 1,089 benign samples for validation, the highest performance on 10-fold validation reported an accuracy of 99.96\%, a precision of 96.34\%, a recall of 99.11\%, and an F1-Score of 100\%.

Ravi et al. \cite{ravi2022attention} introduce an attention-based multi-dimensional deep learning framework that processes byte sequences from ELF files, emphasizing Linux operating systems. The proposed model integrates CNN and Bi-LSTM, along with an attention mechanism to enhance feature extraction from extensive byte sequences, addressing both spatial features of byte values and sequential byte information. The architecture features a dual pathway, with one path using CNN with an attention mechanism to extract and focus on spatial features of the byte values, and the other path using Bi-LSTM to learn from both forward and backward sequences, capturing temporal dependencies within the byte sequences. A global attention layer is placed after the CNN and Bi-LSTM layers to bolster the model's capacity to specify the crucial information in lengthy byte sequences. The outcomes from CNN and LSTM are concatenated before being directed through a series of fully connected layers with batch normalization and dropout to prevent overfitting and ensure generalization. Evaluated on the CDMC-2020-IoMT-Malware dataset, encompassing 36,236 samples, and further tested for malware classification on the Big-2015 dataset, the model exhibited proficiency in IoMT malware detection with an accuracy, precision, recall, and F1-Score of 95\%. For IoMT malware classification, it achieved an accuracy of 94\%, precision of 96\%, recall of 87\%, and an F1-Score of 91\%.

\textbf{AI-powered and privacy-preserving malware detection in IoMT}. Nawshin et al. \cite{nawshin2024ai} introduce an AI-powered malware detection model named DP-RFECV-FNN, which integrates Differential Privacy (DP) within a Feedforward Neural Network (FNN) for zero trust security, particularly in the IoMT domain. Combining the privacy-preserving capabilities of Differential Privacy with the robust feature selection of Recursive Feature Elimination with Cross-Validation (RFECV) and the classification power of FNN, this model ensures both high security and privacy. The model is trained on data containing instances of malware from the CCCS-CIC-AndMal-2020 dataset for static features and the CICMalDroid 2020 dataset for dynamic features, covering various malware types such as Adware, SMS, Ransomware, Riskware, and Trojan. The feature representation involves extracting static features like activities, intents, services, permissions, and metadata from AndroidManifest.xml files, as well as dynamic features such as system calls, binder calls, and composite behaviors from .apk files, processed through extraction, selection, and normalization steps. For the static feature dataset, the model achieved an accuracy of up to 99.21\%, and for the dynamic feature dataset, it demonstrated an accuracy of up to 94.36\%.

\textbf{DRL and DNN-based hybrid models for malware detection in IoMT}. He et al. \cite{he2023image} introduce a hybrid deep learning model named DRL-HMD, which utilizes Deep Neural Networks (DNN) and Deep Reinforcement Learning (DRL) for zero-day malware detection in IoMT devices. By combining DNN models with a DRL-based decision-making agent, the model ensures fast response and improved accuracy by dynamically selecting the best malware detector at runtime. The model is trained on data containing various types of malware, including worms, viruses, botnets, ransomware, spyware, adware, trojans, rootkits, and backdoors, sourced from VirusShare and VirusTotal repositories. The feature representation involves converting hardware performance counter (HPC) data into small-size images using a 2D embedding algorithm. These images are processed through transfer learning to enhance the DNN models. For the datasets used, the DRL-HMD model achieved a high detection rate of 99\% in both F1-score and AUC, with a false positive rate of only 0.01\% and a false negative rate of 1\%.

\subsection{\textbf{Internet of Vehicles (IoV)}}



Few works have addressed malware analysis, particularly targeting IoVs. We summarize these handful of works in Table~\ref{tab:iov}.

\textbf{Hybrid deep learning models for intrusion detection in IoV}. Safi Ullah et al. \cite{ullah2022hdl} introduce a hybrid deep learning model named HDL-IDS, which utilizes LSTM and GRU networks for intrusion detection in the IoV. By combining LSTM and GRU models with a ReLU-activated dense layer, the model provides a fast response and leverages the deep layers of LSTM and GRU for superior results. The model is trained on data containing instances of cyber-attacks from a combined DDoS dataset (CIC DoS CI-CIDS 2017 and CSE-CIC-IDS 2018) and a car-hacking dataset, which cover instances of DDoS attacks, fuzzing, and spoofing within vehicular networks. The feature representation involves extracting numerical values from the datasets, which are then processed and normalized. For the DDoS dataset, the HDL-IDS model achieved an accuracy of 99.5\%, precision of 99.5\%, recall of 99.6\%, and an F1-Score of 99.5\%. For the car-hacking dataset, it demonstrated even higher performance metrics, with an accuracy of 99.9\%, precision of 99.9\%, recall of 99.9\%, and an F1-Score of 99.9\%.

\textbf{RNN-based approaches for detecting vehicle cyber-physical intrusions}. Loukas et al. \cite{loukas2017cloud} employed a RNN with LSTM layers to analyze time-series data for detecting vehicle cyber-physical intrusions. This methodology efficiently captures the temporal dynamics associated with various intrusion types by processing a rich set of features extracted from the vehicle’s operational data. These features include network traffic rates (incoming and outgoing), CPU utilization, disk writing rates, physical vibrations (measured by accelerometers), power consumption, and wheel rotation speeds (monitored through magnetic encoders). The study targeted robotic vehicles operating on the Fedora Linux system and compared the RNN's performance against a Multi-Layer Perceptron (MLP) and traditional machine learning techniques, including Logistic Regression, Decision Trees, Random Forest, and Support Vector Machines (SVM). The RNN model demonstrated the best accuracy, outperforming others with 95.4\% on DoS datasets, 83.2\% on command injection scenarios, and 82.2\% on malware attacks. Furthermore, the research explored computational offloading to enhance intrusion detection capabilities in resource-constrained environments. By offloading the computational tasks to external infrastructure, such as cloud services, the system managed to handle resource constraints more effectively. This approach allowed for real-time data processing and timely intrusion detection, despite the limited processing power available onboard the vehicle.

\textbf{CNN-based models for intrusion detection in IoV}. Nie et al. \cite{nie2020data} introduce a deep learning model named IDS-CNN, which utilizes a CNN for intrusion detection in the IoV. The CNN model is designed to analyze the link load behaviors of Road Side Units (RSUs) against various attacks, capturing network traffic's temporal and spatio-temporal features. The model is trained on data containing cyber-attack instances generated from a testbed imitating an IoV scenario with RSUs and OBUs and attackers using LOIC to implement DDoS attacks. The IDS-CNN model achieved accuracy, precision, recall, and F1-score all over 97\% for both scenarios with 12 and 30 OBUs. However, the model performs better with shorter attack lengths, showing higher accuracy with 2 timeslots compared to 7 timeslots.

\textbf{LSTM-based autoencoders for detecting anomalous events in ITS}. Ashraf et al. \cite{ashraf2020novel} utilize an LSTM autoencoder for detecting anomalous events in Intelligent Transportation Systems (ITS). Combining the LSTM autoencoder with advanced statistical feature extraction techniques ensures fast response times and improved accuracy in detecting both known and unknown (zero-day) attacks. The model is trained on data containing cyber-attack instances from a car hacking dataset and the UNSW-NB15 dataset, covering various attack types on in-vehicle communications and external network communications. The feature representation involves statistical features extracted from network traffic data, such as packet counts, message sizes, and traffic metrics. For the car hacking dataset, the model achieved an accuracy of over 99\%, and for the UNSW-NB15 dataset, it achieved an accuracy of 96\%.

\textbf{Deep CNN models for intrusion detection in in-vehicle networks}. Song et al. \cite{song2020vehicle} introduce a deep learning model named DCNN-IDS, which utilizes a reduced Inception-ResNet architecture for intrusion detection in in-vehicle networks. By combining inception and residual blocks with a frame builder module, the model is capable of learning temporal sequential patterns in CAN bus data. The model is trained on data containing cyber-attack instances from a real vehicle's CAN bus, including DoS attacks, fuzzy attacks, gear spoofing, and RPM spoofing. The feature representation involves transforming bit-stream data into 29 × 29 bitwise frames. For the RPM spoofing dataset, the DCNN-IDS model achieved the best performance with a precision of 99.99\%, recall of 99.94\%, F1-score of 99.96\%, minimum False Negative Rate (FNR) of 7\%, and an Error rate of 4\%.

\section{\textbf{Open Challenges and Future Research Directions}}
\label{sec:Open Challenges and Future Research Directions}

While deep learning is achieving significant results and shaping the field of malware analysis in the XIoT domain, several open research challenges still need to be addressed. Tackling these issues requires multifaceted approaches that consider the constraints and vulnerabilities of XIoT systems. This section discusses several open challenges and highlights potential future directions.


\subsubsection{\textbf{Lack of Studies}}
There is a notable lack of research studies focusing on XIoT malware detection using deep learning techniques, especially for IoV and IoBT. XIoT-specific malware research is necessary because these domains face unique challenges such as high mobility, dynamic topology, the integration of heterogeneous devices, and the need to analyze spatio-temporal data.

Unlike traditional IoT environments, which often have static or limited mobility, XIoT domains like IoV involve vehicles that are constantly moving and communicating with each other and with infrastructure, creating a highly dynamic and complex network environment. Similarly, IoBT encompasses a wide range of devices used in battlefield scenarios, necessitating robust and resilient security measures due to the critical nature of military operations and the high likelihood of targeted attacks by adversarial entities.

Moreover, the analysis of spatio-temporal data, which involves understanding the spatial and temporal relationships between data points, is crucial in XIoT. This adds another layer of complexity to malware detection efforts, underscoring the need for specialized research to effectively address these unique challenges. 

\subsubsection{ \textbf{5G Integration Risks in XIoT}}
Integrating 5G technology in XIoT domains significantly enhances operational efficiency but also introduces substantial security and data privacy risks. The increased functional complexity of these systems—including various interconnected devices such as sensors, actuators, intelligent appliances, and connected vehicles—creates new attack vectors. These interconnected devices often have varying security measures, processing power, and resources, making them attractive targets for malware attacks \cite{ahmed2022multilayer}.

The integration of 5G expands the network's reach, connecting more devices and thereby increasing the number of potential entry points for cyber-attacks. Each connected device represents a possible vulnerability that can be exploited. With more devices transmitting data over the network, the risk of sensitive information being intercepted by malicious actors also increases. Furthermore, the high-speed data transfer capabilities of 5G can be leveraged to quickly exfiltrate large volumes of data, increasing the potential damage of cyber-attacks.

\subsubsection{\textbf{Resource Constraints and Leveraging Edge and Fog Computing in XIoT Devices}}
XIoT devices typically have limited computational, communication, and processing resources, making it challenging to implement advanced deep learning models directly on these devices due to their high computational and storage requirements. This limitation hampers the achievement of efficient real-time malware detection \cite{kim2022iiot,esmaeili2022iiot,shah2023heucrip,amin2022deep}. Deep learning techniques generally demand substantial resource utilization, such as RAM, and can be sluggish during the training phase, which is not ideal for real-time processes or XIoT domains \cite{amin2022deep}. Additionally, deploying deep learning models, particularly those with complex architectures, necessitates significant computational resources, limiting their feasibility in resource-constrained XIoT environments \cite{esmaeili2022iiot,shah2023heucrip,kim2022iiot}. Ensuring robustness against adversarial attacks, such as through Deep Eigenspace Learning techniques to mitigate junk code insertion \cite{azmoodeh2018robust}, often requires additional computational resources, further compounding the challenge in XIoT domains.

To address these challenges, utilizing edge and fog computing can help offload the processing load from resource-constrained XIoT devices \cite{loukas2017cloud}. However, several issues remain that require further research and development. In real offloading scenarios, both computing and communication resources are limited. Mobile devices and edge servers might not have sufficient computational power or bandwidth to handle all offloaded tasks, leading to delays in malware detection. Varying wireless channel conditions can affect the performance of offloading, with poor channel conditions resulting in increased latency and reduced reliability of offloaded tasks. Moreover, malware detection tasks often have strict latency requirements, making it challenging to ensure that offloaded tasks are completed within the required time frame, especially when multiple users are involved, and the network is congested. Furthermore, while offloading computation-intensive tasks to edge servers can save energy on mobile devices, it may lead to significant energy consumption at the edge servers. Balancing energy consumption between mobile devices and edge servers is a critical challenge that needs to be addressed to ensure efficient and effective malware detection in XIoT domains.

\subsubsection{ \textbf{High Rate of Malware Attacks on Android Operating System}} 
The Android OS, widely used in both smartphones and XIoT devices, is particularly vulnerable to malware attacks due to its large user base, openness, and ease of application development. The presence of millions of apps, including those from third-party stores, creates numerous vulnerabilities that are frequently exploited by malware. The high volume of malicious applications and the constant emergence of new threats pose significant challenges in maintaining robust security measures \cite{amin2022deep}.

\subsubsection{\textbf{Optimizing Deep Learning Models for Resource-Constrained XIoT Devices}}
Despite the use of deep learning techniques for feature extraction and enhanced detection accuracy, continuously adapting to new malware variants and maintaining high detection performance in XIoT domains remains a significant challenge. XIoT devices, due to their diverse and heterogeneous nature, are particularly vulnerable to sophisticated and evolving malware threats. The continuous evolution of malware necessitates frequent updates to detection models, which can be resource-intensive and difficult to deploy across a wide range of XIoT devices \cite{khan2021hybrid}.

To address these challenges, continuous learning frameworks that can adapt to new malware variants in real-time are essential. Future research could focus on integrating online learning techniques and reinforcement learning to ensure malware detection systems remain effective as threats evolve. These advanced learning methods can enable models to continuously update and improve their detection capabilities without the need for extensive retraining, making them more suitable for the dynamic and resource-constrained XIoT environments \cite{wu2018enhancing}.

Additionally, research can explore lightweight and energy-efficient deep learning models that can be deployed directly on XIoT devices. Techniques such as model pruning, quantization, and novel architectures like TinyML \cite{TensorFlowWebsite, coffen2021tinydl} can be investigated to reduce computational and storage requirements without sacrificing detection accuracy. Another promising solution is leveraging federated learning, which distributes computational tasks across multiple devices, thus reducing the load on any single device. This approach enables the training of robust models without overloading resource-constrained XIoT devices, ensuring efficient and effective malware detection.

\subsubsection{\textbf{Efficient Feature Extraction from Unstructured and Spatio-Temporal Data}}
Malware detection in XIoT domains involves processing not only unstructured data, such as byte sequences from ELF files, but also spatio-temporal data generated in the IoV domain \cite{nie2020data}. Extracting meaningful features from these data types is challenging due to their lack of structure, noise, and complex relationships between spatial and temporal information. The deep learning models designed to handle spatio-temporal data require significant computational resources for training, posing a challenge in XIoT domains where devices often have limited computational capabilities. Additionally, unstructured data, such as raw network traffic data, lacks a predefined format, making it difficult to process and analyze. Sophisticated algorithms and extensive preprocessing are required to extract meaningful features, which can be resource-intensive and time-consuming.

Investigating new methods for processing and extracting features from unstructured and spatio-temporal data can significantly improve detection accuracy in XIoT domains. Future research could explore advanced techniques such as federated learning, which allows for decentralized training of models across multiple devices, thereby preserving privacy and reducing the need for central data storage. Natural Language Processing (NLP) techniques can be adapted to handle raw byte sequences and network traffic data, providing a more structured approach to feature extraction. Self-supervised learning, which leverages the vast amounts of unlabeled data typical in XIoT, can improve model accuracy by learning useful representations without requiring extensive labeled datasets.

\subsubsection{\textbf{Integration of Explainable AI (XAI) in Malware Detection}}
Embedding explainable AI techniques in malware detection systems can enhance transparency and understanding of the produced results \cite{galli2024explainability}. This is particularly crucial in XIoT domains, such as IoV and IoBT, where the complexity and heterogeneity of devices increase the difficulty of understanding and verifying the actions of deep learning models. In these scenarios, the ability to interpret model decisions can help quickly identify and address security threats, ensuring that critical operations are not disrupted by false positives or false negatives.

Moreover, explainable AI can assist in gaining the trust of users and operators in XIoT domains by providing clear insights into the decision-making processes of the models. This transparency can lead to better risk management and compliance with stringent security requirements, ultimately enhancing the reliability and effectiveness of malware detection systems in these environments.

\section{\textbf{CONCLUSION}}
\label{sec:conclusion}

The rapid development of IoT devices and their integration into various sectors—such as industrial, medical, automotive, and military—has significantly increased the demand for robust security mechanisms. Traditional malware detection methods have proven increasingly ineffective against sophisticated IoT malware, thus requiring the adoption of advanced deep learning-based techniques. This survey has comprehensively explored the application of deep learning techniques for malware detection across the XIoT domains, including IoT, IIoT, IoMT, IoV, and IoBT.

Our findings underscore the efficiency of deep learning techniques in identifying malware. These techniques automatically extract intricate patterns from vast datasets, thereby significantly reducing manual labor and achieving high detection and classification accuracy.

However, several challenges still hinder XIoT malware detection. A primary challenge is the heterogeneity of XIoT devices, which include a broad range of hardware architectures and operating systems, complicating the implementation of uniform security measures. Additionally, the resource-constrained nature of many XIoT devices limits the feasibility of deploying computationally intensive deep learning models. Privacy concerns also arise from the extensive data collection required to train these models, posing significant security challenges.

Future research should prioritize the development of lightweight and efficient deep learning models specifically designed for resource-constrained XIoT devices. Additionally, exploring federated learning approaches could enhance privacy by enabling models to be trained locally on devices, thus avoiding the transfer of sensitive data to centralized servers.

\bibliographystyle{IEEEtran}
\bibliography{bibfile}
\end{document}